\documentclass[letterpaper,twocolumn,10pt]{article}

\usepackage{dblfloatfix}
\usepackage{graphicx}
\usepackage{textcomp}
\usepackage{xcolor}
\usepackage{fancyhdr}
\usepackage{subcaption}
\usepackage{comment}
\usepackage{pifont}
\usepackage{relsize}
\usepackage{cases}
\usepackage{tabularx}
\usepackage[utf8]{inputenc}
\usepackage{kotex}
\usepackage{booktabs}
\usepackage{multirow}
\usepackage{array}
\usepackage{graphbox}
\usepackage{circledsteps}
\usepackage{setspace}
\usepackage{times}
\usepackage{lipsum}
\usepackage{epstopdf}
\usepackage{pythonhighlight}
\usepackage{colortbl}
\usepackage{enumitem}
\usepackage{flushend}

\usepackage{kotex}
\usepackage[normalem]{ulem}
\usepackage{xcolor}
\usepackage{soul}
\usepackage{xspace}
\usepackage{lipsum}

\definecolor{codegray}{rgb}{0.5,0.5,0.5}
\lstdefinestyle{pythonStyle}{
  basicstyle=\tiny\ttfamily\footnotesize,
  commentstyle=\color{codegray},
  frame=single,
  language=Python,
  stepnumber=1,
  numbers=left,
  numbersep=5pt,
  numberstyle=\tiny\color{codegray},
  tabsize=2,
  showspaces=false,
  showstringspaces=false,
  mathescape,
  moredelim=**[is][\color{red}]{~}{~},
  moredelim=**[is][\color{blue}]{<}{>},
  moredelim=**[is][\color{orange}]{@}{@},
  literate={\\~}{{\textasciitilde}}1
  {\\<}{{\unichar{"003C}}}1
  {\\>}{{\unichar{"003E}}}1
  {\\@}{{\unichar{"0040}}}1
}

\newcommand{\mycomment}[1]{}

\newcommand\highlight[1][yellow]{%
  \bgroup
  \markoverwith{\textcolor{#1}{\vrule width.2em height.8em depth.2em}}%
  \ULon
}

\newcommand{\todo}[1]{\color{white}\textbf{\highlight[black]{TODO: [#1]}}\color{black}\xspace}
\newcommand{\fixme}[1]{\color{red}\textbf{\highlight[yellow]{FIXME: [#1]}}\color{black}\xspace}

\newcommand{\pointer}[1]{\color{white}\textbf{\highlight[red]{POINTER: [#1 is working here]}}\color{black}\xspace}

\newcommand{\explain}[1]{\textbf{\highlight[yellow]{[#1]}}\xspace}
\newcommand{\hlist}[1]{\textbf{\highlight[yellow]{#1}}\xspace}

\usepackage{usenix-2020-09}
  \newcommand{\circledblack}[1]{\Circled[fill color=black, inner color=white]{\scriptsize{#1}}}

\usepackage[framemethod=tikz]{mdframed}
\usepackage[numbers,sort&compress]{natbib}
\tikzset{
    vertical align/.style={
        baseline=-.5*(height("$+$")-depth("$+$"))
    }
}
\newcommand*{\squareNum}[1]{\tikz[baseline=(char.base)]{
            \node[shape=rectangle,fill,inner sep=1.8pt] (char) {\textcolor{white}{\scriptsize{#1}}};}}
\newcommand*{\circleNum}[1]{\tikz[baseline=(char.base)]{
              \node[shape=circle,fill,inner sep=0.9pt] (char) {\textcolor{white}{\scriptsize{#1}}};}}
\makeatletter
\def\hlinewd#1{%
\noalign{\ifnum0=`}\fi\hrule \@height #1 %
\futurelet\reserved@a\@xhline}
\makeatother

\usepackage[hang,flushmargin]{footmisc}
\newcommand\blfootnote[1]{%
  \begingroup
  \renewcommand\thefootnote{}\footnote{#1}%
  \addtocounter{footnote}{-1}%
  \endgroup
}

\definecolor{darkred}{HTML}{C00000}
\makeatletter
\newcommand{\thickhline}{
    \noalign {\ifnum 0=`}\fi \hrule height 3pt
    \futurelet \reserved@a \@xhline
}
\newcolumntype{"}{@{\hskip\tabcolsep\vrule width 1.2pt}}
\makeatother
\begin{document}

\date{}

\title{\Large \bf
    Hardware/Software Co-Programmable Framework for Computational SSDs to Accelerate Deep Learning Service on Large-Scale Graphs
}

\author{
{\rm Miryeong Kwon \hspace{15pt} Donghyun Gouk \hspace{15pt} Sangwon Lee \hspace{15pt} Myoungsoo Jung}\\
Computer Architecture and Memory Systems Laboratory\\
Korea Advanced Institute of Science and Technology (KAIST)\\
http://camelab.org
} 

\maketitle
\blfootnote{This paper has been accepted at the 20th USENIX Conference on File and Storage Technologies (FAST '22). This material is presented to ensure timely dissemination of scholarly and technical work.}
\begin{abstract}
Graph neural networks (GNNs) process large-scale graphs consisting of a hundred billion edges. In contrast to traditional deep learning, unique behaviors of the emerging GNNs are engaged with a large set of graphs and embedding data on storage, which exhibits complex and irregular preprocessing.

We propose a novel deep learning framework on large graphs, \textit{HolisticGNN}, that provides an easy-to-use, near-storage inference infrastructure for fast, energy-efficient GNN processing.
To achieve the best end-to-end latency and high energy efficiency, HolisticGNN allows users to implement various GNN algorithms and directly executes them where the actual data exist in a holistic manner. It also enables RPC over PCIe such that the users can simply program GNNs through a graph semantic library without any knowledge of the underlying hardware or storage configurations.

We fabricate HolisticGNN's hardware RTL and implement its software on an FPGA-based computational SSD (CSSD).
Our empirical evaluations show that the inference time of HolisticGNN outperforms GNN inference services using high-performance modern GPUs by 7.1$\times$ while reducing energy consumption by 33.2$\times$, on average.
\end{abstract}

\section{Introduction}
\label{sec:intro}
Graph neural networks (GNNs) have recently emerged as a representative approach for learning graphs, point clouds, and manifolds.
Compared to traditional graph analytic methods, GNNs exhibit much higher accuracy in a variety of prediction tasks \cite{xu2018powerful,hamilton2017representation,scarselli2008graph,yao2019graph,zhao2019t,li2019semi}, and their generality across different types of graphs and algorithms allows GNNs to be applied by a broad range of applications such as social networks, knowledge graphs, molecular structure, and recommendation systems \cite{fan2019graph,lin2020kgnn,chen2019graph,he2020lightgcn}.
The state-of-the-art GNN models such as GraphSAGE \cite{hamilton2017inductive} further advance to infer unseen nodes or entire new graphs by generalizing geometric deep learning (DL).
The modern GNN models in practice sample a set of subgraphs and DL feature vectors (called \emph{embeddings}) from the target graph information, and aggregate the sampled embeddings for inductive node inferences \cite{hamilton2017inductive,ying2018graph}.
This \emph{node sampling} can significantly reduce the amount of data to process, which can decrease the computation complexity to infer the results without an accuracy loss \cite{chen2018fastgcn,hamilton2017inductive,you2018graph}.

While these node sampling and prior model-level efforts for large graphs make the inference time reasonable, GNNs yet face system-level challenges to improve their performance.
First, GNNs experience a completely different end-to-end inference scenario compared to conventional DL algorithms.
In contrast to the traditional DLs, GNNs need to deal with real-world graphs consisting of billions of edges and node embeddings \cite{wang2018billion,eksombatchai2018pixie}.
The graph information (graph and node embeddings) initially reside in storage and are regularly updated as raw-format data owing to their large size and persistence requirements.
As GNNs need to understand the structural geometry and feature information of given graph(s), the raw-format data should be loaded into working memory and reformatted in the form of an adjacency list before the actual inference services begin.
These activities take a significant amount of time since the graph information often exceeds hundreds of GBs or even a TB of storage \cite{wilkening2021recssd,eisenman2018bandana}.
We observe that the pure inference latency, that all the previous GNN studies try to optimize, accounts for only 2\% of the end-to-end inference service time when we execute diverse GNN models in a parallel system employing high-performance GPUs \cite{gtx1060,rtx3090} and an SSD \cite{intelp4600}.
We will analyze this performance bottleneck issue with detailed information in Section \ref{subsec:challenge}.

Second, GNNs consist of various computing components, which are non-trivial to fully accelerate or parallelize over conventional computing hardware.
As GNNs are inspired by conventional DL algorithms such as convolution neural networks and representative learning \cite{wu2020comprehensive,scarselli2008graph,hamilton2017inductive}, several data processing parts of GNN computing are associated with dense matrix computing.
While these matrix multiplications can be accelerated by existing data processing units (DPUs) such as systolic architectures, the graph-natured operations of GNNs can neither be optimized with DPU's multiplication hardware nor with GPUs' massive computing power \cite{auten2020hardware}.

A promising alternative to address the aforementioned challenges is employing in-storage processing (ISP) to serve GNN inferences directly from the underlying storage.
While ISP is very a well-known solution heavily studied in the literature for the past few decades \cite{jo2016yoursql,koo2017summarizer,seshadri2014willow,gu2016biscuit,lee2014accelerating,quero2015self}, it has unfortunately not been widely adopted in real-world systems \cite{balasubramonian2014near}.
There are several reasons, but the most critical issue of ISP is ironically its concept itself, which co-locates flash and computing unit(s) into the same storage box.
As flash is not a working memory but a block device, it is integrated into the storage box with complicated firmware and multiple controllers \cite{jung2011architecture,zhang2020scalable, colgrove2015purity}.
These built-in firmware and controllers are not easily usable for the computation that users want to offload as it raises many serious technical challenges such as programmability, data protection, and vendor dependency issues.
In this work, we advocate a new concept of \emph{computational SSD} (CSSD) architectures that locate reconfigurable hardware (FPGA) near storage in the same PCIe subsystem \cite{sniacssd, xilinxcssd, smartssd}.
In contrast to ISP, CSSD can maximize peer-to-peer acceleration capability and make it independent from specific storage firmware and controller technologies.
However, it is challenging to configure everything that users want to program and/or download in the form of completely full hardware logic into FPGA from scratch.

We propose \textit{HolisticGNN}, a hardware and software co-programmable framework that leverages CSSD to accelerate GNN inference services near storage.
HolisticGNN offers a set of software and hardware infrastructures that execute GNN tasks where data exist and infer the results from storage in a holistic manner.
Generally speaking, the software part of HolisticGNN enables users to program various GNN algorithms and infer embedding(s) directly atop the graph data without the understanding complexities of the underlying hardware and device interfaces.
On the other hand, our hardware framework provides fundamental hardware logic to make CSSD fully programmable.
It also provides an architectural environment that can accelerate various types of GNN inferences with different hardware configurations.


For fast and energy-efficient GNN processing, our framework is specifically composed of three distinguishable components: i) graph-centric archiving system (\emph{GraphStore}), ii) programmable inference client and server model (\emph{GraphRunner}), and iii) accelerator building system (\emph{XBuilder}).
The main purpose of GraphStore is to bridge the semantic gap between the graph abstraction and its storage representation while minimizing the overhead of preprocessing.
GraphStore manages the user data as a graph structure rather than exposing it directly as files without any intervention of host-side software.
This allows diverse node sampling and GNN algorithms to process the input data near storage immediately. GraphStore also supports efficient mutable graph processing by reducing the SSD's write amplification.

To accommodate a wide spectrum of GNN models, it is necessary to have an easy-to-use, programmer-friendly interface.
GraphRunner processes a series of GNN inference tasks from the beginning to the end by allowing users to program the tasks using a computational graph.
The users can then simply transfer the computational graph into the CSSD and manage its execution through a remote procedure call (RPC).
This does not require cross-compilation or storage stack modification to program/run a user-defined GNN model.
We enable RPC by leveraging the traditional PCIe interface rather than having an extra hardware module for the network service, which can cover a broad spectrum of emerging GNN model implementations and executions in CSSD.

On the other hand, XBuilder manages the FPGA hardware infrastructure and accelerates diverse GNN algorithm executions near storage.
It first divides the FPGA logic die into two regions, \emph{Shell} and \emph{User}, using the dynamic function exchange (DFX) technique \cite{ug909}.
XBuilder then secures hardware logic necessary to run GraphStore and GraphRunner at Shell while placing DL accelerator(s) to User.
The Shell and User hardware are programmed to CSSD as two separate bitstreams, such that we can reprogram the User with a different bitstream at any time.
To this end, XBuilder implements a hardware engine in Shell by using an internal configuration access port, which downloads a bitstream and programs it to User.




We implement HolisticGNN on our CSSD prototype that places a 14$nm$ FPGA chip \cite{virtexup} and 4TB NVMe device \cite{intelp4600} under a same PCIe switch.
We also prototype the software framework of HolisticGNN on the CSSD bare-metal, and we fabricate/test various GNN accelerator candidates within CSSD, such as a many-core processor, systolic arrays, and a heterogeneous (systolic+vector) processor.
Our evaluations show that the inference time of HolisticGNN outperforms GNN inference services using high-performance GPUs by 7.1$\times$ while consuming 33.2$\times$ less energy, on average.

\section{Background}
\label{sec:background}
\begin{figure}
  \centering
  \includegraphics[width=\linewidth]{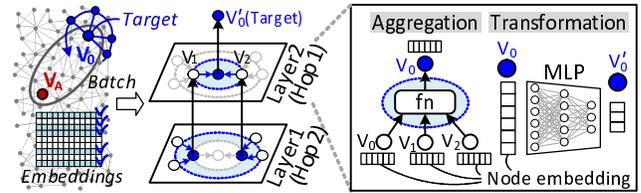}
  \begin{subfigure}{\linewidth}
    \centering
    \setlength\tabcolsep{0pt}
    \renewcommand*{\arraystretch}{0.3}
    \begin{tabularx}{\textwidth}{
            p{\dimexpr.27\linewidth-2\tabcolsep-1.3333\arrayrulewidth}
            p{\dimexpr.33\linewidth-2\tabcolsep-1.3333\arrayrulewidth}
            p{\dimexpr.39\linewidth-2\tabcolsep-1.3333\arrayrulewidth}
        }
           \vspace{-9pt} \caption{Preprocessing.}\label{fig:bck_gnn_algo1}
         & \vspace{-9pt} \caption{GNN processing.}\label{fig:bck_gnn_algo2}
         & \vspace{-9pt} \caption{Layer's details.}\label{fig:bck_gnn_algo3}
    \end{tabularx}
  \end{subfigure}
  \vspace{-27pt}
  \caption{Overview of basic GNN algorithm.}
  \label{fig:bck_gnn_algo}
  \vspace{-18pt}
\end{figure}

\subsection{Graph Neural Networks}\label{subsec:gnn}
Graph neural networks (GNNs) generalize conventional DL to understand structural information in the graph data by incorporating feature vectors (\emph{embeddings}) in the learning algorithms \cite{hamilton2017representation,hamilton2017inductive,ying2018graph}.
GNNs can capture topological structures of the local neighborhood (per node) in parallel with a distribution of the neighborhood's node embeddings \cite{chen2018fastgcn,hamilton2017inductive,you2018graph,battaglia2018relational}.

\noindent \textbf{General concept.}
As shown in Figure \ref{fig:bck_gnn_algo}, GNNs in general take three inputs, a graph, the corresponding node embeddings (e.g., user profile features), a set of unseen/seen nodes to infer, called \emph{batch}.
Since the internal memory of GPUs is insufficient to accommodate all the inputs, it is essential to reduce the size of the graph and embeddings by preprocessing them appropriately (Figure \ref{fig:bck_gnn_algo1}), which will be explained in Section \ref{subsec:datapre}.
GNNs then analyze the preprocessed structural information with node embeddings over multiple computational layers (Figure \ref{fig:bck_gnn_algo2}).
Each layer of GNNs is composed of two primary execution phases, called neighborhood \textit{aggregation} and node \textit{transformation} \cite{wu2020comprehensive, xu2018powerful}, which are all performed for neighbors at different hop distances (connected to a target node in the batch).
Specifically, as shown in Figure \ref{fig:bck_gnn_algo3}, the aggregation is a simple function to accumulate node embeddings of the target node's neighbors, whereas the transformation converts the aggregated results to a new node embedding using one or more traditional \emph{multi-layer perceptrons} (MLPs \cite{hornik1989multilayer,hornik1991approximation}).
Therefore, the aggregation processes data relying on graph structures and mainly exhibits irregular, graph-natured execution patterns.
In contrast, the transformation computing procedure is very similar to that of conventional neural networks (e.g., CNNs and RNNs), but it does not require heavy computation.
For example, GNNs mostly use only 2$\sim$3 layers \cite{defferrard2016convolutional,kipf2016semi,xu2018powerful,wang2019neural,velivckovic2017graph}, whereas Google BERT employs more than 24 layers and needs to perform heavy matrix multiplications \cite{devlin2018bert}.

Note that, while the massive parallel computing of GPUs is very well-optimized for many DL algorithm executions, these characteristics of GNNs (e.g., irregular execution pattern and relatively lightweight computation) allow other processing architectures to be a better fit for GNN acceleration.

\noindent \textbf{Model variations.}
Based on how to aggregate/transform embeddings, there is a set of variant GNNs, but \emph{graph convolution network} (GCN \cite{kipf2016semi}), \emph{graph isomorphism network} (GIN \cite{xu2018powerful}), and \emph{neural graph collaborative filtering} (NGCF \cite{wang2019neural}) are the most popular GNN models used in node/graph classification and recommendation systems \cite{yao2019graph,zhao2019t,li2019semi,you2018graph,fan2019graph,he2020lightgcn}.

\begin{figure}
  \centering
  \includegraphics[width=\linewidth]{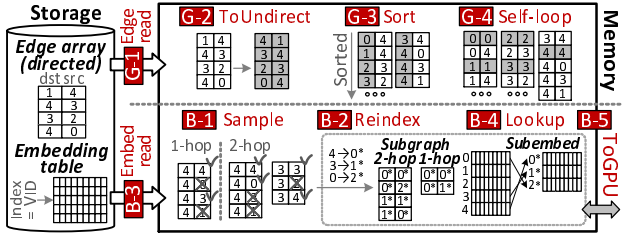}
  \vspace{-20pt}
  \caption{Holistic viewpoint of GNN computing (\textbf{G}: Graph preprocessing, \textbf{B}: Batch preprocessing).}
  \label{fig:bck_gnn_pre}
  \vspace{-17pt}
\end{figure}

Specifically, GCN uses an \emph{``average-based aggregation''} that normalizes the embeddings by considering the degree of neighbor nodes.
This prevents cases where a specific embedding has excessively large features, thereby losing other embeddings that have relatively small amounts of data in the aggregation phase.
In contrast, GIN uses a \emph{``summation-based aggregation''} that does not normalize the embeddings of both the target node (self-loop) and its neighbors.
In addition, GIN gives a learnable self-weight to the target node embedding to avoid unexpectedly losing its feature information due to the heavy states and features of the target node's neighbors.
To precisely capture the structural characteristics of the given graph, GIN uses a two-layer MLP structure, making the combination more expressively powerful.
GCN and GIN suppose that all the feature vectors of a given graph have the same level of weight, which are widely used for node and graph classifications \cite{yao2019graph,marcheggiani2017encoding,zhao2019t}.
Instead of using a simple average or summation for the aggregation, NGCF takes the similarity among the given graph's embeddings into account by applying an element-wise product to neighbors' embeddings.

Even though there are several variants of GNNs, they all require the graph's geometric information to analyze embeddings during the aggregation and transformation phases. Thus, it is necessary for GNNs to have an easy-to-access, efficient graph and embedding data structures.

\subsection{Graph Dataset Preprocessing}\label{subsec:datapre}
The graph data offered by a de-facto graph library such as SNAP \cite{leskovec2016snap} in practice deal with edge information in the form of a text file.
The raw graph file includes an (unsorted) edge array, each being represented by a pair of destination and source \emph{vertex identifiers} (VIDs).
Most GNN frameworks such as deep graph library (DGL) \cite{wang2019dgl} and pytorch geometric (PyG) \cite{fey2019pyg} preprocess the graph dataset to secure such easy-to-access graph and embeddings as a VID-indexed table or tensor.
In this work, we classify these graph dataset preprocessing tasks into two: i) \emph{graph preprocessing} and ii) \emph{batch preprocessing}.
While graph preprocessing is required only for the geometrical data (including the initialization), batch preprocessing should be performed for each inference service.

\noindent \textbf{Graph preprocessing.}
Since the majority of emerging GNN algorithms are developed based on spatial or spectral networks encoding \emph{``undirected''} geometric characteristics \cite{defferrard2016convolutional,kipf2016semi}, the main goal of this graph preprocessing is to obtain a sorted, undirected graph dataset.
As shown in the top of Figure \ref{fig:bck_gnn_pre}, it first loads the edge array (raw graph) from the underlying storage to the working memory [$\squareNum{G-1}$].
To convert the array to an undirected graph, the GNN frameworks (e.g., DGL) allocate a new array and copy the data from the edge array to the new array by swapping the destination and source VIDs for each entry (\{\textit{dst}, \textit{src}\}$\rightarrow$\{\textit{src}, \textit{dst}\}) [$\squareNum{G-2}$].
The frameworks merge and sort the undirected graph, which turns it into a VID-indexed graph structure [$\squareNum{G-3}$].
As the target node to infer is also included in the 1-hop neighborhood, the frameworks inject self-loop information (an edge connecting a vertex to itself) to the undirected graph as well (\{0,0\}, \{1,1\}, $\cdots$ \{4,4\}) [$\squareNum{G-4}$]. If there is no self-loop information, the aggregation of GNNs cannot reflect a visiting node's features, which in turn reduces the inference accuracy significantly.


\noindent \textbf{Batch preprocessing.}
Large-scale real-world graphs consist of a hundred billion edges, and each node of the edges is associated with its own embedding containing thousands of DL features.
The number of nodes and the embedding size that the current GNN models process are typically an order of magnitude greater than heavy featured DL applications, such as natural language processing \cite{wilkening2021recssd,eisenman2018bandana}.
Thus, for a given batch, the frameworks in practice perform \emph{node sampling} such as random walk \cite{yang2019knightking} and unique neighbor sampling \cite{hamilton2017inductive}. The node sampling specifically extracts a set of subgraphs and the corresponding embeddings from the original (undirected) graph datasets before aggregating and transforming the feature vectors, which can significantly reduce data processing pressures and decrease the computing complexity without an accuracy loss \cite{hamilton2017inductive,jangda2021accelerating}.
Since the sampled graph should also be self-contained, the subgraphs and embeddings should be reindexed and restructured. We refer to this series of operations as \emph{batch preprocessing}.

The bottom of Figure \ref{fig:bck_gnn_pre} shows an example of batch preprocessing.
For the sake of brevity, this example assumes that the batch includes a just single target, $V_4$ (VID=4), the given sampling size is 2, and GNN is modeled with two layers (two hops).
This example first reads all $V_4$'s neighbors and extracts two nodes from the undirected graph (in a random manner) [$\squareNum{B-1}$].
This generates a subgraph including the 1-hop neighbors, which is used for GNN's layer 2 computation (L2).
For the sampled nodes ($V_4$ and $V_3$), it reads their neighbor nodes (2-hop) and samples the neighborhood again for GNN's layer 1 (L1).
Since the number of nodes has been significantly reduced, the GNN frameworks allocate new VIDs in the order of sampled nodes (4$\rightarrow$ 0*, 3$\rightarrow$ 1*, and 0$\rightarrow$ 2*) and create L1 and L2 subgraphs for 2-hop and 1-hop neighbors, respectively [$\squareNum{B-2}$].
It then composes an embedding table whose index is the VID of each sampled node.
To this end, the frameworks first need to load the embeddings from the underlying storage to working memory [$\squareNum{B-3}$], called global embeddings, and lookup the embeddings of L1's subgraph ($V_4$, $V_0$, and $V_3$) [$\squareNum{B-4}$].
Lastly, the subgraphs and sampled embedding table are required to transfer from the working memory to the target GPU's internal memory [$\squareNum{B-5}$]. 

\subsection{Challenge Analysis}
\label{subsec:challenge}
\begin{figure}
  \captionsetup[figure]{aboveskip=-15pt,belowskip=-15pt}
  \captionsetup[subfigure]{aboveskip=0pt}
  \begin{subfigure}{0.56\linewidth}
      \includegraphics[width=1\linewidth]{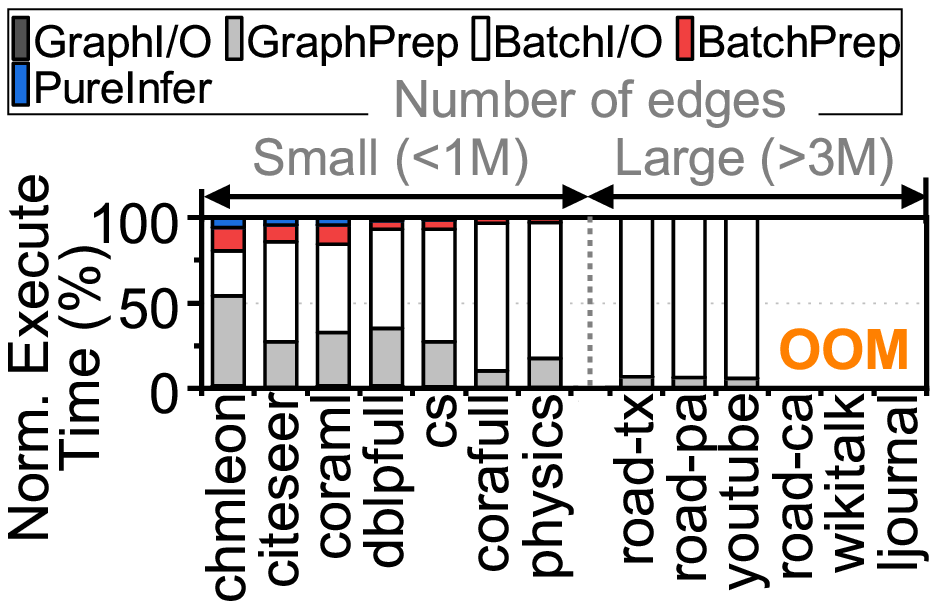}
      \caption{Latency breakdown.}
      \label{fig:motiv1}
  \end{subfigure}
  \begin{subfigure}{0.43\linewidth}
      \includegraphics[width=1\linewidth]{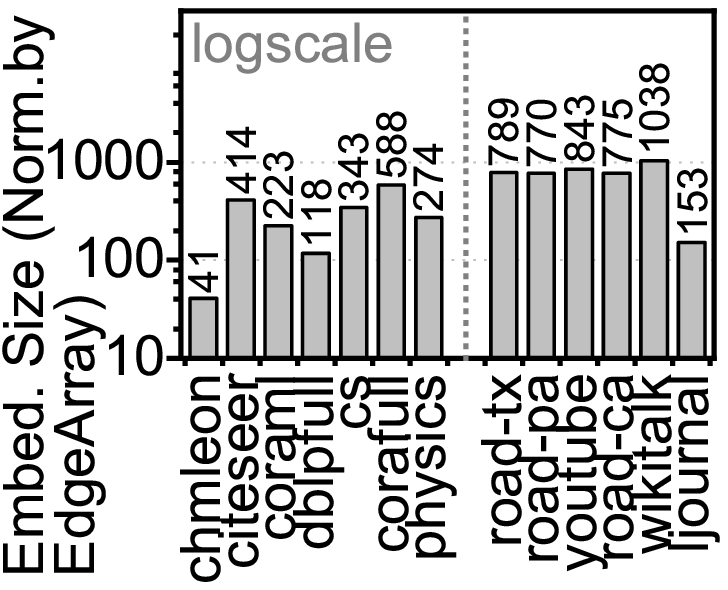}
      \caption{Embed vs. Edge array.}
      \label{fig:motiv2}
  \end{subfigure}
  \vspace{-10pt}
  \caption{End-to-End GNN execution.}
  \label{fig:motiv}
  \vspace{-20pt}
\end{figure}

While there is less system-level attention on the management of graph and batch preprocessing, their tasks introduce heavy storage accesses and frequent memory operations across the boundary of user space and storage stack.
To be precise, we decompose the \emph{``end-to-end GCN inference''} times across 14 real-world graph workloads (coming from \cite{yang2016revisiting,shchur2018pitfalls,rozemberczki2021multi,leskovec2016snap}) into the latency of graph preprocessing (\texttt{GraphPrep}), batch preprocessing (\texttt{BatchPrep}), GCN inference processing (\texttt{PureInfer}), and storage accesses for graph (\texttt{GraphI/O}) and embeddings (\texttt{BatchI/O}).
Since the storage access latency being overlapped with the latency of preprocessing computation is invisible to users, this breakdown analysis excludes such latency, and the results are shown in Figure \ref{fig:motiv1}.
The detailed information of the evaluation environment is provided by Section \ref{sec:evaluation}.
One can observe from this breakdown analysis that \texttt{PureInfer} only takes 2\% of the end-to-end inference latency, on average.
Specifically, \texttt{BatchI/O} accounts for 61\% of the most end-to-end latency for the small graphs having less than 1 million edges.
Before the sorted and undirected graph is ready for batch preprocessing, \texttt{BatchI/O} cannot be processed.
Since \texttt{GraphPrep} includes a set of heavy (general) computing processes such as a radix sort, \texttt{GraphPrep} also consumes 28\% of the end-to-end latency for these small graphs.
As the graph size increases (> 3 million edges), \texttt{BatchI/O} becomes the dominant contributor of the end-to-end GNN inference time (94\%, on average).
Note that the inference system has unfortunately stopped the service during the preprocessing due to out-of-memory (OOM) when it handles large-scale graphs (>3 million edges) such as \texttt{road-ca}, \texttt{wikitalk}, and \texttt{ljournal}.
This OOM issue can be simply addressed if one services the GNN inference from where the data exist.
In addition, the heavy storage accesses and relatively lightweight computing associated with inference (\texttt{PureInfer}) make adoption of the in-storage processing concept \cite{6558444} reasonable to shorten the end-to-end inference latency.

Figure \ref{fig:motiv2} normalizes the size of the embedding table to that of the edge array (graph) across all the graphs that we tested.
As shown in this comparison, embedding tables of the small and large graphs are greater than the edge arrays for those graphs by 285.7$\times$ and 728.1$\times$, on average, respectively.
This is because an embedding has thousands of DL features, each represented using floating-point values with high precision \cite{ying2018graph,rozemberczki2021multi}.
In contrast, an entry of the edge arrays contains only a simple integer value (VID).
This characteristic makes batching preprocessing I/O intensive while inducing graph preprocessing to be computation-intensive.
\section{Storage as a GNN Accelerator}
\label{sec:design-overview}
In-storage processing (ISP) is well-studied in the research literature \cite{keeton1998case,acharya1998active,seshadri2014willow,lee2014accelerating,quero2015self,gu2016biscuit,koo2017summarizer,jun2018grafboost,ruan2019insider}, but it has been applied to accelerate limited applications such as compression and key-value management in real-world systems.
There are several reasons, but the greatest weakness of ISP ironically is that it needs to process data where data is stored, i.e., at the flash device.
Flash cannot be directly used as block storage because of its low-level characteristics, such as I/O operation asymmetry and low reliability \cite{murugan2011rejuvenator,jung2012taking,choi2017exploiting,wu2012delta,pan2011exploiting,jung2019design}.
Thus, flash requires tight integration with multiple firmware and controller modules \cite{zhang2018flashshare,zhang2020scalable}, which renders ISP difficult to be implemented within an SSD.

\begin{figure}
  \centering
  \captionsetup[figure]{aboveskip=-15pt,belowskip=-15pt}
  \captionsetup[subfigure]{aboveskip=0pt}
  \includegraphics[width=1\linewidth]{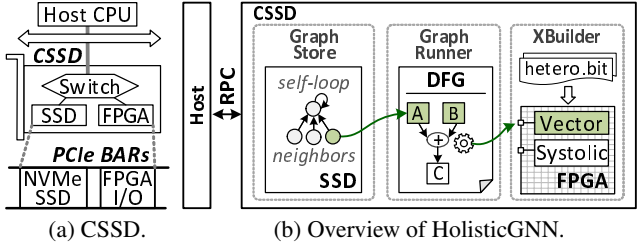}
  \begin{subfigure}{.3\linewidth}
    \vspace{-15pt} \caption{CSSD.} \label{fig:over_cssd1}
  \end{subfigure}
  \begin{subfigure}{.69\linewidth}
    \vspace{-15pt} \caption{Overview of HolisticGNN.} \label{fig:over_cssd2}
  \end{subfigure}
  \vspace{-26pt}
  \caption{Enabling CSSD for near storage GNN processing.\label{fig:over_cssd}}
  \vspace{-19pt}
\end{figure}

In contrast to ISP, as shown in Figure \ref{fig:over_cssd1}, the new concept of computational SSDs (CSSDs) decouples the computing unit from the storage resources by locating \emph{reconfigurable hardware} (FPGA) near SSD in the same PCIe subsystem (card) \cite{smartssd}.
CSSD allows the hardware logic fabricated in FPGA to access the internal SSD via the internal PCIe switch.
To this end, the host is responsible for writing/reading data on the SSD using the I/O region of NVMe protocol while giving the data's block address to the FPGA through its own I/O region, whose address is designated by PCIe's base address register \cite{ug1382}.
While CSSD is promising to realize near-data processing \cite{sniacssd,kwon2021fast}, it is non-trivial to automate all end-to-end procedures of GNN inference over hardware-only logic because of the variety of GNN model executions.
For example, the aggregation and/or combination of GNNs can be accelerated with parallel hardware architecture, but GNN's graph traversing, dataset preprocessing, and embedding handling are impractical to be programmed into hardware because of their graph-natured computing irregularities.

\subsection{Overview of HolisticGNN}
HolisticGNN is a hardware and software co-programmable framework that leverages CSSD to accelerate the end-to-end GNN inference services near storage efficiently.
The software part of our framework offers easy-to-use programming/management interfaces and performs GNN preprocessing directly from where the data is stored, thereby minimizing the aforementioned storage access overhead. 
HolisticGNN can also eliminate the out-of-memory issue for deep learning on large-scale graphs.
On the other hand, our framework's hardware logic and administration module provide a low-overhead bare-metal computing environment and reconfigurable hardware to accelerate GNN model executions.

Figure \ref{fig:over_cssd2} illustrates a high-level view of HolisticGNN, which is composed of three major modules: i) graph-centric archiving system (\emph{GraphStore}), ii) programmable inference model (\emph{GraphRunner}), and iii) accelerator builder (\emph{XBuilder}).
Generally speaking, GraphStore prepares the target graph data and embeddings in a ready-to-access structure that the tasks of batch preprocessing can immediately use without preprocessing the datasets.
On the other hand, GraphRunner executes a series of GNN inference tasks from the beginning to the end, and it processes the graph datasets by directly accessing SSD through GraphStore.
GraphRunner also provides a \emph{dataflow graph} (DFG) based program and execution model to support easy-to-use and flexible implementation of a wide spectrum of GNN algorithms.
This enables the users to simply generate a DFG and deliver it to HolisticGNN, which can dynamically change the end-to-end GNN inference services without cross-compilation and/or understanding underlying hardware configurations.
Lastly, XBuilder makes CSSD simply reconfigurable and has heterogeneous hardware components to satisfy the diverse needs of GNN inference acceleration services.
XBuilder also provides several kernel building blocks, which abstract the heterogeneous hardware components. This can decouple a specific hardware acceleration from the GNN algorithm implementation.

Each module of our framework exposes a set of APIs through remote procedure calls (RPCs) to users.
These APIs are not related to GNN programming or inference services, but to framework management such as updating graphs, inferring features, and reprogramming hardware logic.
Since CSSD has no network interface for the RPC-based communication, we also provide an \emph{RPC-over-PCIe} (RoP) interface that overrides the conventional PCIe to enable RPC between a host and CSSD without an interface modification.

\begin{table}
  \centering
  \resizebox{\linewidth}{!}{%
  \begin{tabular}{c|l"c|l}
    \hlinewd{1.2pt}
    \textbf{Service type} & \multicolumn{1}{c"}{\textbf{RPC function}} & \hspace{0.5em}\textbf{Service type} & \multicolumn{1}{c}{\textbf{RPC function}} \\ \hline
    \begin{tabular}[c]{@{}c@{}}GraphStore\\ (Bulk)\end{tabular}
      & \begin{tabular}[c]{@{}l@{}}\texttt{UpdateGraph}\\ \texttt{(EdgeArray, Embeddings)}\end{tabular}
      & \hspace{0.5em}\multirow{2}{*}{\begin{tabular}[c]{@{}c@{}}GraphStore\\ (Unit, Get)\end{tabular}}
      & \texttt{GetEmbed(VID)}
        \\ \cline{1-2} \cline{4-4}

    \multirow{5}{*}{\begin{tabular}[c]{@{}c@{}}GraphStore\\ (Unit, Update)\end{tabular}}
      & \texttt{AddVertex(VID,Embed)}
      &
      & \texttt{GetNeighbors(VID)}
        \\ \cline{2-4}

      & \texttt{DeleteVertex(VID)}
      & \hspace{0.5em} \multirow{2}{*}{\begin{tabular}[c]{@{}c@{}}Graph\\ Runner\end{tabular}}
      & \texttt{Run(DFG, batch)}
        \\ \cline{2-2} \cline{4-4}

      & \texttt{AddEdge(dstVID,srcVID)}
      &
      & \texttt{Plugin(shared\_lib)}
        \\ \cline{2-4}

      & \texttt{DeleteEdge(dstVID,srcVID)}
      & \hspace{0.5em} \multirow{2}{*}{XBuilder}
      & \multirow{2}{*}{\texttt{Program(bitfile)}}
        \\ \cline{2-2}

      & \texttt{UpdateEmbed(VID,Embed)}
      &
      &
        \\ \hlinewd{1.2pt}
    \end{tabular}}
    \vspace{-8pt}
  \caption{RPC services of HolisticGNN.}
  \vspace{-20pt}
  \label{tbl:rpc}
\end{table}

\subsection{Module Decomposition}
\noindent \textbf{Graph-centric archiving system.}
The main goal of GraphStore is to bridge the semantic gap between graph and storage data without having a storage stack.
As shown in Table \ref{tbl:rpc}, GraphStore offers two-way methods for the graph management, \emph{bulk operations} and \emph{unit operations}.
The bulk operations allow users to update the graph and embeddings with a text form of edge array and embedding list.
For the bulk operations, GraphStore converts the incoming edge array to an adjacency list in parallel with transferring the embedding table, and it stores them to the internal SSD.
This makes the conversion and computing latency overlapped with the heavy embedding table updates, which can deliver the maximum bandwidth of the internal storage.
In contrast, the unit operations of GraphStore deal with individual insertions (\texttt{AddVertex()}/\texttt{AddEdge()}), deletions (\texttt{DeleteVertex()}/\texttt{DeleteEdge()}), and queries (\texttt{GetEmbed()}/\texttt{GetNeighbors()}) for the management of graph datasets.
When GraphStore converts the graph to storage semantic, it uses VID to \emph{logical page number} (LPN) mapping information by being aware of a long-tailed distribution of graph degree as well as flash page access granularity. The LPNs are used for accessing CSSD's internal storage through NVMe, which can minimize the write amplification caused by I/O access granularity differences when CSSD processes GNN services directly on the SSD. The design and implementation details are explained in Section \ref{subsec:gs_detail}.

\noindent \textbf{Programmable inference model.}
GraphRunner decouples CSSD task definitions from their actual implementations, which are called \emph{C-operation} and \emph{C-kernel}, respectively.
To program a GNN model and its end-to-end service, the users can write a DFG and download/execute to CSSD by calling GraphRunner's RPC interface (\texttt{Run()}) with a request batch containing one or more target nodes.
Figure \ref{fig:dfg} shows a simple example of GCN implementation.
The DFG has a set of input nodes for the target sampled subgraphs, embeddings, and weights, which are connected to a series of C-operations such as averaging features (Mean), matrix multiplication (Matmul), a non-linear function (ReLU), and output feature vector (Out\_embedding).
This DFG is converted to a computational structure by sorting the node (C-operation) and edge (input node information) in topological order.
Once the DFG is downloaded through HolisticGNN's RoP serialization, GraphRunner's engine deserializes it and executes each node with appropriate inputs by checking the registered C-operations and C-kernels in CSSD.
The users may want to register more C-operations/kernels because of adoption of a new GNN model or hardware logic.
To meet the requirement, GraphRunner offers a Plugin mechanism registering a pair of C-operation/C-kernel and a new device configuration as a shared object.
We will explain the details of GraphRunner in Section \ref{subsec:swframe}.


\begin{table}
  \centering
  \resizebox{\linewidth}{!}{%
  \begin{tabular}{c|l"c|l}
    \hlinewd{1.2pt}
    \textbf{API type} & \multicolumn{1}{c"}{\textbf{Function}} & \hspace{0.5em}\textbf{API type} & \multicolumn{1}{c}{\textbf{Operation format}} \\ \hline
    \multirow{4}{*}{\begin{tabular}[c]{@{}c@{}}DFG\\ Creation\end{tabular}}
      & \texttt{createIn(name)}
      & \hspace{0.5em}\multirow{6}{*}{XBuilder}
      & \texttt{GEMM(inputs, output)}
        \\ \cline{2-2} \cline{4-4}

      & \texttt{createOp(name)}
      &
      & \texttt{ElementWise(inputs, output)}
        \\ \cline{2-2} \cline{4-4}

      & \texttt{createOut(name)}
      &
      & \texttt{Reduce(inputs, output)}
        \\ \cline{2-2} \cline{4-4}

      & \texttt{save(graph)}
      &
      & \texttt{SpMM(inputs, output)}
        \\ \cline{1-2} \cline{4-4}

    \multirow{2}{*}{Plugin}
      & \texttt{RegisterDevice(newDevice)}
      &
      & \multirow{2}{*}{\texttt{SDDMM(inputs, output)}}
        \\ \cline{2-2}

      & \texttt{RegisterOpDefinition(newOp)}
      &
      &
        \\ \hlinewd{1.2pt}
    \end{tabular}}
  \vspace{-8pt}
  \caption{Programming interface of HolisticGNN.}
  \label{tbl:api}
  \vspace{-20pt}
\end{table}

\noindent \textbf{Accelerator builder.}
To make the FPGA of CSSD easy to use, we configure CSSD's hardware logic die into two groups, \emph{Shell} and \emph{User} logic, by leveraging a dynamic function exchange (DFX) mechanism \cite{ug909}.
DFX allows hardware to be modified blocks of logic with separate \emph{bitfiles} that contain the programming information for an FPGA.
XBuilder secures Shell logic associated with irregular tasks of GNN management, including GraphStore and GraphRunner executions, while managing User logic for users to reprogram the hardware in accelerating GNN algorithms via XBuilder's RPC interface (\texttt{Program()}).
\texttt{Program()} moves a partial bitfile into the internal memory and asks an XBuilder engine to reconfigure User logic hardware using the bitfile via FPGA internal configuration access port \cite{ug974,ug570}.

XBuilder also abstracts the registered device (at User logic) by providing a set of basic building blocks to the users as shown in Table \ref{tbl:api}.
The building blocks basically implement what DL and GNN algorithms mostly use, such as general matrix multiplication (GEMM) and sparse matrix multiplication (SpMM), across different legacy acceleration hardware such as multi-core, vector processor, systolic architecture.
XBuilder's building blocks operate specific hardware based on the device priority designated by C-kernel that user defines.
We will discuss this in details in Section \ref{subsec:reconfig}.

\subsection{Enabling RPC over PCIe}
While the key method to program GNN models (and request their inference services) is associated with DFG, the underpinning of such a device-to-device communication method is RPC.
As the investigation of efficient RPC is not the purpose of this work, we use Google's gRPC \cite{grpc} and implement our RPC-based interfaces themselves (e.g., \texttt{UpdateGraph()}, \texttt{Run()}, etc.) using interface definition language (IDL) \cite{protobuf}.
We also modify the gRPC stack to enable RPC services without changing hardware and storage interfaces.

\begin{figure}
  \centering
  \includegraphics[width=1\linewidth,bb=0 8 247 86]{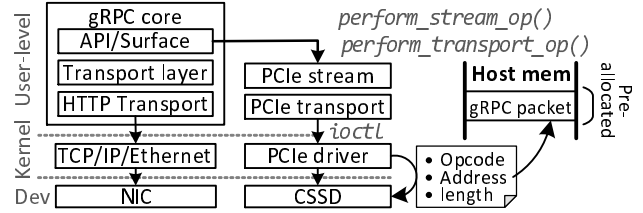}
  \vspace{-16pt} \caption{RPC over PCIe (RoP).\label{fig:rop}}
  \vspace{-18pt}
\end{figure}

Figure \ref{fig:rop} explains the gRPC stack and how HolisticGNN enables gRPC over PCIe.
The host-side gPRC interfaces are served by a user-level gRPC core, which manages transport and HTTP connection.
We place two gRPC plugin interfaces (\texttt{perform\_stream\_op()} and \texttt{perform\_transport\_op()}), each forwarding the requests of gRPC core's transport layer and HTTP transport to our PCIe stream and PCIe transport modules.
Specifically, the PCIe stream is responsible for managing \texttt{stream} data structures, which are used for gPRC packet handling.
Similarly, the PCIe transport deals with the host and CSSD connection by allocating/releasing \texttt{transport} structures.
While the original gPRC core is built upon a kernel-level network stack including TCP/IP and Ethernet drivers, we place a PCIe kernel driver connected to the PCIe transport.
It supports gRPC's send/receive packet services and other channel establishment operations to the PCIe transport module via \texttt{ioctl}.
The PCIe kernel driver also provides preallocated buffer memory to the PCIe stream through a memory-mapped I/O (\texttt{mmap}).
This buffer memory contains gPRC packet's metadata and message such that the PCIe driver lets the underlying CSSD know the buffer's location and offset.
Specifically, the PCIe drive prepares a PCIe command that includes an opcode (send/receive), address (of memory-mapped buffer), and length (of the buffer).
When the driver writes the command to FPGA's designated PCIe memory address, CSSD parses the command and copies the data from the memory-mapped buffer into FPGA-side internal memory for gRPC services.

\section{Design Details and Implementation}
\label{sec:design-archive}
\subsection{Efficient Storage Accesses for Graphs}
\label{subsec:gs_detail}
\begin{figure}[]
  \begin{subfigure}{\linewidth}
    \centering
    \includegraphics[width=1\linewidth]{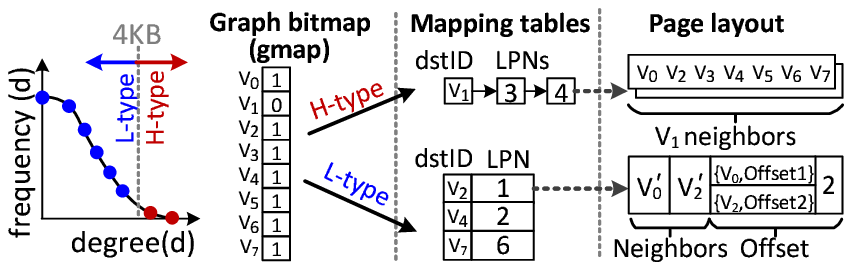}
    \begin{tabularx}{\textwidth}{
      p{\dimexpr.3\linewidth-2\tabcolsep-1.3333\arrayrulewidth}
      p{\dimexpr.7\linewidth-2\tabcolsep-1.3333\arrayrulewidth}
      }
        \vspace{-14pt} \caption{Distribution.} \label{fig:gs01}
      & \vspace{-14pt} \caption{Two types of mapping.} \label{fig:gs02}
    \end{tabularx}
  \end{subfigure}
  \vspace{-28pt}
  \caption{GraphStore's mapping structure.}
  \label{fig:gs0}
  \vspace{-20pt}
\end{figure}

GraphStore maintains the graph datasets as an adjacency list and an embedding table to handle the geometric information and feature vectors.
While the embedding table is stored in sequential order (and thus it does not require page-level mapping), the adjacency list is maintained in two different ways by considering the efficiency of graph searches/updates: i) high-degree graph mapping (\emph{H-type}) and ii) low-degree graph mapping (\emph{L-type}).
As shown in Figure \ref{fig:gs01}, the power-law graph's natures make a few nodes have severely heavy neighbor nodes (high-degree) \cite{reittu2004power}. These high-degree nodes account for a small fraction of the entire graph, but they have a high potential to be frequently accessed and updated (because of their many neighbors).
H-type mapping is therefore designed towards handling the graph's long-tailed distribution well, while L-type mapping is structured to achieve high efficiency of flash page management.

\noindent \textbf{Mapping structure.}
As shown in Figure \ref{fig:gs02}, GraphStore has a graph bitmap (\emph{gmap}), which explains what kind of tables are used for mapping (per VID).
Basically, the mapping entry for both types of mapping tables pairs a VID and an LPN (VID-to-LPN), but the corresponding page stores different data with its own page layout.
The H-type page maintains many neighbors' VID in a page, and its mapping table entry indicates a linked list in cases where the neighbors of the target (source) VID cannot be stored in a flash page (4KB). 
The L-type page also contains many VIDs, but their source VIDs vary.
To this end, the end of page has meta-information that indicates how many nodes are stored and where each node exists on the target page (offset).
Thus, L-type mapping table's VID is the biggest VID among VIDs stored in the corresponding page.

\begin{figure}[]
  \begin{subfigure}{\linewidth}
    \centering
    \includegraphics[width=1\linewidth]{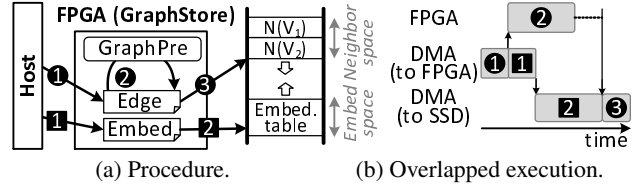}
    \begin{tabularx}{\textwidth}{
        p{\dimexpr.5\linewidth-2\tabcolsep-1.3333\arrayrulewidth}
        p{\dimexpr.5\linewidth-2\tabcolsep-1.3333\arrayrulewidth}
        }
          \vspace{-14pt} \caption{Procedure.} \label{fig:gs11}
        & \vspace{-14pt} \caption{Overlapped execution.} \label{fig:gs12}
    \end{tabularx}
  \end{subfigure}
  \vspace{-28pt} \caption{Bulk operations.} \label{fig:gs1}
  \vspace{-20pt}
\end{figure}

\noindent \textbf{Bulk operation.}
As shown in Figure \ref{fig:gs1}, while the embedding table is stored from the end of LPN (embedding space), the graph pages are recorded from the beginning of storage (neighbor space), similar to what the conventional memory system stack does.
Note that, the actual size of graph(s) is small enough, but it is involved in heavy graph preprocessing, and the majority of graph datasets are related to their node embeddings (cf. Section \ref{subsec:challenge}).
Thus, when an edge array (graph) arrives, GraphStore performs graph preprocessing and flushes pages for the graph, but it does not immediately update them to the target storage.
Instead, GraphStore begins to write the embedding table into the embedding space in a sequential manner while preprocessing the graph.
This can make heavy storage accesses (associated with the embeddings) entirely overlap with the computation burst of graph preprocessing (associated with adjacency list conversions).
From the user's viewpoint, the latency of bulk operation is the same as that of data transfers and embedding table writes.

\noindent \textbf{Unit operations.}
GraphStore's unit operations support mutable graph management corresponding to individual vertex/edge updates or queries.
Figure \ref{fig:gs2} shows how to find out neighbors of $V_4$ and $V_5$, each being classified as high-degree and low-degree nodes.
In this example, as the gmap indicates that $V_4$ is managed by the H-type mapping, the neighbors can be simply retrieved by searching where the target VID is.
In contrast, the page managed by L-type contains many neighborhoods each being associated with different VID. Therefore, when GraphStore searches the mapping table, it considers the range of VIDs, stored in each entry.
For example, $V_5$ is within the range of $V_4$ and $V_6$, GraphStore first needs to retrieve the page corresponding $V_6$.
It finds out $V_5$'s offset and the next VID's offset ($V_6$) by considering the number of node counts in the page's meta-information, which indicates the data chunk containing $V_5$'s neighbors.

\begin{figure}[b]
  \vspace{-22pt}
  \begin{subfigure}{\linewidth}
    \centering
    \includegraphics[width=1\linewidth,bb=0 5 197 65]{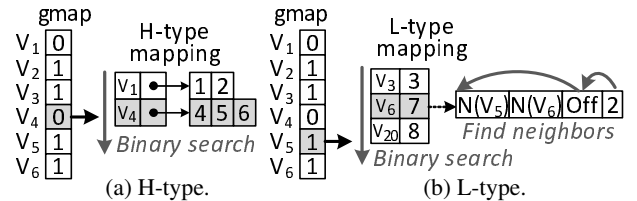}
    \begin{tabularx}{\textwidth}{
      p{\dimexpr.5\linewidth-2\tabcolsep-1.3333\arrayrulewidth}
      p{\dimexpr.5\linewidth-2\tabcolsep-1.3333\arrayrulewidth}
      }
        \vspace{-12pt} \caption{H-type.} \label{fig:gs21}
      & \vspace{-12pt} \caption{L-type.} \label{fig:gs22}
    \end{tabularx}
  \end{subfigure}
  \vspace{-28pt}
  \caption{Unit operations (Get).}
  \label{fig:gs2}
  \vspace{-5pt}
\end{figure}

\begin{figure*}[b]
  \vspace{-15pt}
  \addtocounter{figure}{1}
  \begin{subfigure}{\linewidth}
    \centering
      \includegraphics[width=1\linewidth]{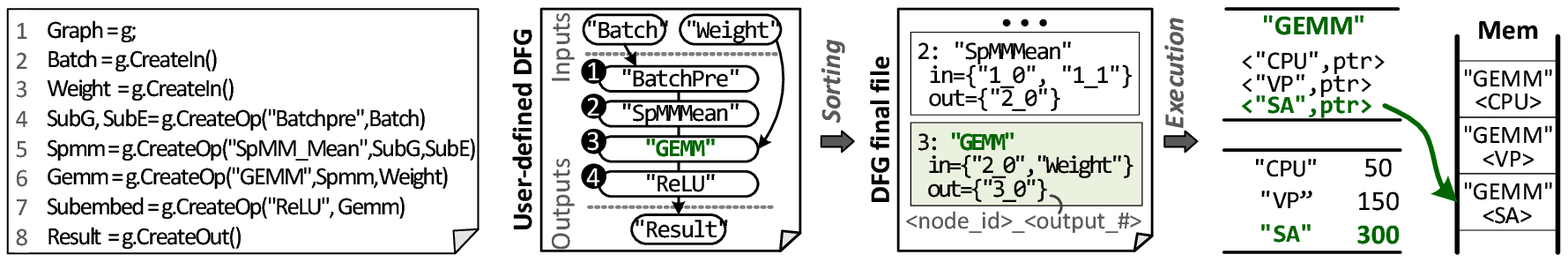}
      \begin{tabularx}{\textwidth}{
          p{\dimexpr.31\linewidth-2\tabcolsep-1.3333\arrayrulewidth}
          p{\dimexpr.24\linewidth-2\tabcolsep-1.3333\arrayrulewidth}
          p{\dimexpr.19\linewidth-2\tabcolsep-1.3333\arrayrulewidth}
          p{\dimexpr.25\linewidth-2\tabcolsep-1.3333\arrayrulewidth}
          }
            \vspace{-10pt} \caption{Example of DFG programming.} \label{fig:graphrunner2}
          & \vspace{-10pt} \caption{Example of DFG.} \label{fig:dfg}
          & \vspace{-10pt} \caption{DFG file generation.} \label{fig:graphrunner3}
          & \vspace{-10pt} \caption{Execution.} \label{fig:graphrunner4}
      \end{tabularx}
  \end{subfigure}
  \vspace{-28pt}
  \caption{Overview of reconfigurable software framework (GraphRunner).}
  \label{fig:graphrunner}
  \vspace{-5pt}
  \addtocounter{figure}{-1}
\end{figure*}

\begin{figure}[]
  \addtocounter{figure}{-1}
  \begin{subfigure}{\linewidth}
    \centering
    \includegraphics[width=1\linewidth]{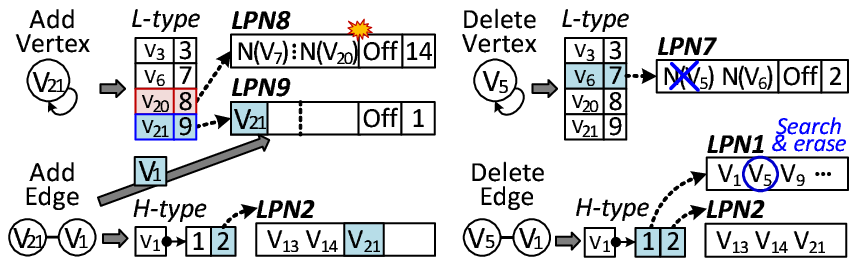}
    \begin{tabularx}{\textwidth}{
      p{\dimexpr.51\linewidth-2\tabcolsep-1.3333\arrayrulewidth}
      p{\dimexpr.49\linewidth-2\tabcolsep-1.3333\arrayrulewidth}
      }
        \vspace{-11pt} \caption{AddEdge/AddVertex.} \label{fig:graphstore1}
      & \vspace{-11pt} \caption{DeleteEdge/DeleteVertex.} \label{fig:graphstore2}
    \end{tabularx}
  \end{subfigure}
  \vspace{-28pt}
  \caption{Unit operations (Update).}
  \label{fig:graphstore}
  \vspace{-16pt}
  \addtocounter{figure}{1}
\end{figure}

Figures \ref{fig:graphstore1} and \ref{fig:graphstore2} show add operations (\texttt{AddEdge()} / \texttt{AddVertex()}) and delete operations (\texttt{DeleteEdge()} / \texttt{DeleteVertex()}).
Let us suppose that $V_{21}$ is given by \texttt{AddVertex()} (Figure \ref{fig:graphstore1}).
GraphStore checks the last entry's page (LPN8) of L-type and tries to insert $V_{21}$ into the page. However, as there is no space in LPN8, GraphStore assigns a new entry ([$V_{21}$,$9$]) to the L-type mapping table by allocating another page, LPN9, and simply appends the vertex information ($V_{21}$) to the page.
Note that, when adding a vertex, it only has the self-loop edge, and thus, it starts from L-type.
When $V_{21}$$\rightarrow$$V_{1}$ is given by \texttt{AddEdge()}, GraphStore makes it an undirected edge ($V_{21}$$\rightarrow$$V_{1}$ \& $V_{21}$$\leftarrow$$V_{1}$).
As $V_{1}$ is H-type, GraphStore checks $V_{1}$'s linked list and places $V_{21}$ to the last page (LPN2). If there is no space in LPN2, it allocates a new page and updates the linked list with the newly allocated page. In contrast, since $V_{21}$ is L-type, GraphStore scans the meta-information of LPN9 and appends $V_{1}$ to the page.
Note that, in cases where there is no space in an L-type page, GraphStore evicts a neighbor set (represented in the page) whose offset of the meta-information is the most significant value.
This eviction allocates a new flash page, copies the neighbor set, and updates L-type mapping table. Since each L-type's destination node has a few source nodes, this eviction case is very rare in practice (lower than 3\% of the total update requests for all graph workloads we tested).

On the other hand, delete operations (Figure \ref{fig:graphstore2}) consist of search and erase tasks.
If \texttt{DeleteVertex()} is called with $V_5$, GraphStore finds out LPN7 and deletes all the neighbors of $V_5$, N($V_5$).
During this time, other neighbors having $V_5$ should also be updated together.
For \texttt{DeleteEdge()} with the given $V_5$$\rightarrow$$V_1$, GraphStore checks all the LPNs indicated by the linked list of $V_1$ and updates the corresponding page (LPN1 in this example) by removing $V_5$. Note that, GraphStore does not have explicit page compaction for the node/edge deletions in an L-type page. This is because, when there is a deletion, GraphStore keeps the deleted VID and reuses it (and the corresponding neighbor set space) for a new node allocation.

\subsection{Reconfiguring Software Framework}
\label{subsec:swframe}
HolisticGNN provides a CSSD library package, which includes the interfaces for C-kernel registration and DFG management as explained previously (cf. Table \ref{tbl:api}).

\noindent \textbf{C-kernel registration and management.}
GraphRunner manages C-kernels by employing a registration mechanism and an execution engine.
GraphRunner has two metadata structures, \emph{Device table} and \emph{Operation table}.
As shown in Table \ref{tab: metadata}, the device table includes currently registered device names and the corresponding priority.
On the other hand, the operation table maintains C-operation names and the address pointers of their C-kernel implementation.
When users implement a C-kernel, it should invoke two registration interface methods of the Plugin library, \texttt{RegisterDevice()} and \texttt{RegisterOpDefinition()}, at its initial time.
\texttt{RegisterDevice()} configures the priority value of the device that users want to execute for any of C-kernels (e.g., ``Vector processor'', 150).
On the other hand, \texttt{RegisterOpDefinition()} registers the device that this C-kernel.
When GraphRunner registers the C-kernel, it places the registration information as a pair of the device name and such C-kernel's pointer.
If there are multiple calls of \texttt{RegisterOpDefinition()} with the same name of a C-operation (but a different name of device), GraphRunner places it in addition to the previously registered C-kernels as a list.
In this example, GraphRunner can recognize that GEMM C-operation defines three C-kernels each using ``CPU'', ``Vector processor'', and ``Systolic array'' by referring to the operation table.
Since the device table indicates ``Systolic array'' has the highest priority, GraphRunner takes the C-kernel associated with ``Systolic array'' for the execution of GEMM C-operation.

\begin{table}[]
  \resizebox{\linewidth}{!}{
    \begin{tabular}{ccccc}
      \multicolumn{2}{c}{\textbf{Device table}}             & \multicolumn{1}{l}{} & \multicolumn{2}{c}{\textbf{Operation table}} \\ \cline{1-2} \cline{4-5}
      \multicolumn{1}{c|}{Name}                 & Priority  &                      & \multicolumn{1}{c|}{Name} & C-kernel \\ \cline{1-2} \cline{4-5}
      \multicolumn{1}{c|}{"CPU"}                &       50  &                      & \multicolumn{1}{c|}{\multirow{3}{*}{"GEMM"}}  & \multirow{3}{*}{
        \begin{tabular}[c]{@{}c@{}}
          \textless{}"CPU", ptr\textgreater \\
          \textless{}"Vector processor", ptr\textgreater \\
          \textless{}"Systolic array", ptr\textgreater{}
        \end{tabular}
      } \\ \cline{1-2}
      \multicolumn{1}{c|}{"Vector processor"}   &       150 &                      & \multicolumn{1}{c|}{}      &       \\ \cline{1-2}
      \multicolumn{1}{c|}{"Systolic array"}     &       300 &                      & \multicolumn{1}{c|}{}      &       \\ \cline{1-2} \cline{4-5}
      \multicolumn{1}{c|}{\dots}                &     \dots &                      & \multicolumn{1}{c|}{\dots} & \dots \\ \cline{1-2} \cline{4-5}
    \end{tabular}}
  \vspace{-10pt}
  \caption{GraphRunner's metadata structure.}
  \label{tab: metadata}
  \vspace{-20pt}
\end{table}

\begin{figure*}[b]
  \addtocounter{figure}{1}
  \centering
  \vspace{-10pt}
  \begin{minipage}[c]{.285\linewidth}
    \includegraphics[width=1\linewidth]{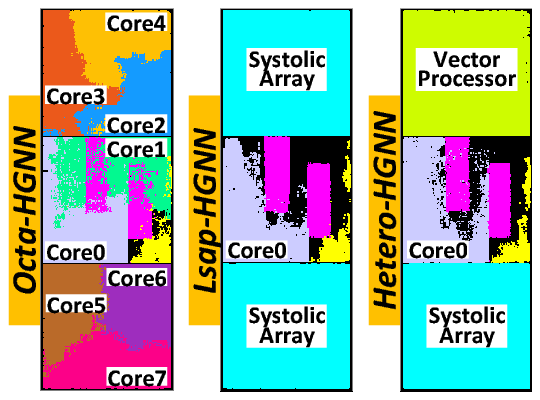}
    \vspace{-18pt}
    \caption{Shell/User prototypes.}
    \label{fig:eval_floorplan}
  \end{minipage}
  \hspace{1pt}
  \begin{minipage}[c]{.16\linewidth}
    \centering
    \includegraphics[align=t,width=1\linewidth]{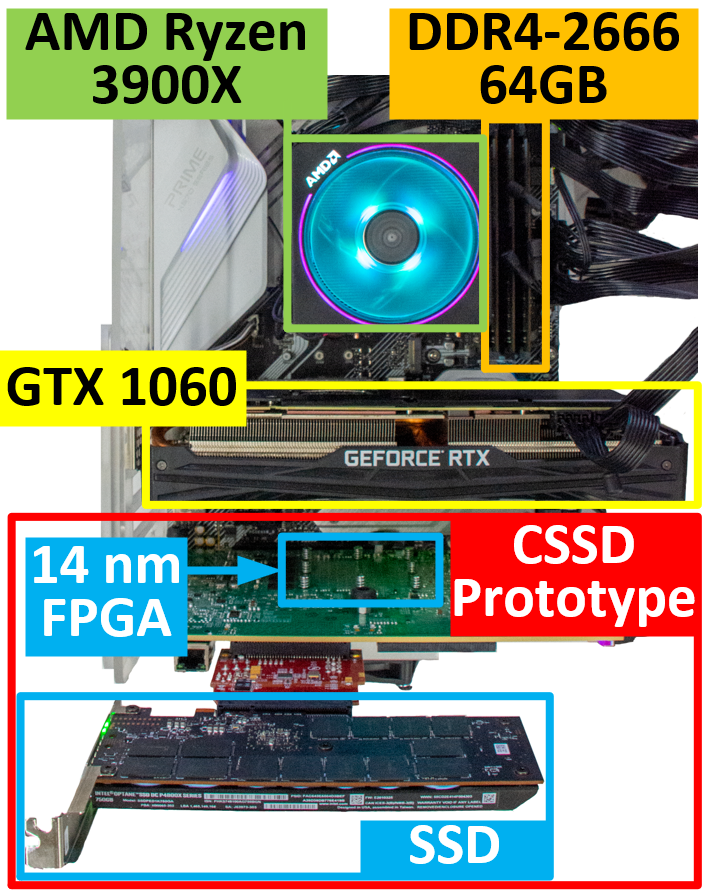}
    \vspace{-9pt}
    \caption{HolisticGNN prototype.}
    \label{fig:hostpc}
  \end{minipage}
  \hspace{1pt}
  \begin{minipage}[c]{.17\linewidth}
    \centering
    \setlength{\tabcolsep}{2pt}
    \renewcommand{\arraystretch}{.8}
    \resizebox{\textwidth}{!}{
      \begin{tabular}{l}
        \toprule
        \multicolumn{1}{c}{\textbf{Host Setup}}           \\ \midrule
          \begin{tabular}{@{}l@{}}
            AMD Ryzen 3900X                                      \\
            \hspace{10pt} {\small 2.2GHz, 12 cores}              \\
            \hspace{10pt} {\small DDR4-2666 16GB x4}             \\
            GTX 1060 6GB \cite{gtx1060}                          \\
            \hspace{10pt} {\small 1.8GHz, 1024 cores (10 SMs)}   \\
            RTX 3090 24GB \cite{rtx3090}                         \\
            \hspace{10pt} {\small 1.74GHz, 10496 cores (82 SMs)} \\
          \end{tabular}                                   \\ \toprule
        \multicolumn{1}{c}{\textbf{FPGA Setup}}           \\ \midrule
          \begin{tabular}{@{}l@{}}
            Xilinx Virtex UltraScale+ \cite{virtexup}   \\
            \hspace{10pt} {\small DDR4-2400 16GB x2}    \\
          \end{tabular}                                   \\  \toprule
        \multicolumn{1}{c}{\textbf{Storage}}              \\  \midrule
          \begin{tabular}{@{}l@{}}
            Intel SSD DC P4600 \cite{intelp4600}    \\
            \hspace{10pt} {\small 3D TLC NAND, 4TB} \\
          \end{tabular} \\  \bottomrule
      \end{tabular}
    }
    \vspace{-9pt}
    \captionof{table}{Host and FPGA setup.}
    \label{tbl:testbeds}
  \end{minipage}
  \hspace{1pt}
  \begin{minipage}[c]{.35\linewidth}
    \centering
    \small
    \setlength{\tabcolsep}{1.5pt}
    \renewcommand{\arraystretch}{.825}
    \resizebox{\linewidth}{!}{%
    \begin{tabular}{@{}llrrrrrr@{}}
      \toprule
      \multicolumn{2}{c}{\multirow{2}{*}{\textbf{Legend}}} & \multicolumn{3}{c}{\textbf{Original Graph}} & \multicolumn{3}{c}{\textbf{Sampled Graph}} \\ \cmidrule(r){3-5} \cmidrule(r){6-8}
        &   & \multicolumn{1}{c}{Vertices}
            & \multicolumn{1}{c}{Edges}
            & \multicolumn{1}{c}{\begin{tabular}[c]{@{}c@{}}Feature\\ Size\end{tabular}}
            & \multicolumn{1}{c}{Vertices}
            & \multicolumn{1}{c}{Edges}
            & \multicolumn{1}{c}{\begin{tabular}[c]{@{}c@{}}Feature\\ Length\end{tabular}} \\ \midrule
      \multicolumn{1}{c}{\multirow{7}{*}{\rotatebox[origin=c]{90}{\begin{tabular}[c]{@{}c@{}}\textbf{Small} \\ \footnotesize (<1M Edges)\end{tabular}}}}
        & chmleon  \cite{rozemberczki2021multi} &  2.3K &    65K &    20 MB & 1,537 &  7,100 & 2326 \\
        & citeseer \cite{yang2016revisiting} &  2.1K &     9K &    29 MB &   667 &  1,590 & 3704 \\
        & coraml   \cite{yang2016revisiting} &  3.0K &    19K &    32 MB & 1,133 &  2,722 & 2880 \\
        & dblpfull \cite{yang2016revisiting} & 17.7K &   123K &   110 MB & 2,208 &  3,784 & 1639 \\
        & cs       \cite{shchur2018pitfalls} & 18.3K &   182K &   475 MB & 3,388 &  6,236 & 6805 \\
        & corafull \cite{yang2016revisiting} & 19.8K &   147K &   657 MB & 2,357 &  4,149 & 8710 \\
        & physics  \cite{shchur2018pitfalls} & 34.5K &   530K & 1,107 MB & 4,926 &  8,662 & 8415 \\ \midrule
      \multicolumn{1}{c}{\multirow{6}{*}{\rotatebox[origin=c]{90}{\begin{tabular}[c]{@{}c@{}}\textbf{Large} \\ \footnotesize (>3M Edges)\end{tabular}}}}
        & road-tx  \cite{leskovec2016snap} & 1.39M &  3.84M &  23.1 GB &   517 &    904 & 4353 \\
        & road-pa  \cite{leskovec2016snap} & 1.09M &  3.08M &  18.1 GB &   580 &  1,010 & 4353 \\
        & youtube  \cite{leskovec2016snap} & 1.16M &  2.99M &  19.2 GB & 1,936 &  2,193 & 4353 \\
        & road-ca  \cite{leskovec2016snap} & 1.97M &  5.53M &  32.7 GB &   575 &    999 & 4353 \\
        & wikitalk \cite{leskovec2016snap} & 2.39M &  5.02M &  39.8 GB & 1,768 &  1,826 & 4353 \\
        & ljournal \cite{leskovec2016snap} & 4.85M & 68.99M &  80.5 GB & 5,756 &  7,423 & 4353 \\ \bottomrule
    \end{tabular}%
    }
    \vspace{-8pt}
    \captionof{table}{Graph dataset characteristics.}
    \label{tbl:dataset}
  \end{minipage}
  \vspace{-5pt}
  \addtocounter{figure}{-1}
\end{figure*}

\noindent \textbf{Handling computational graphs.}
DFG management interfaces of the CSSD library (\texttt{CreateIn()}, \texttt{CreateOut()} and \texttt{CreateOp()}) are used for explaining how C-operations are mapped to DFG's nodes and how their input and output parameters are connected together (Table \ref{tbl:api}).

Figure \ref{fig:graphrunner2} shows how the users can create a DFG to implement a GCN inference service as an example.
The input and output of this DFG is \texttt{Batch}, \texttt{Weight}, and \texttt{Result}.
\texttt{BatchPre} is the first C-operation that takes Batch as its input (\circleNum{1}), and the result is forwarded to \texttt{SpMM\_Mean} C-operation (\circleNum{2}), which performs GCN's average-base aggregation.
Then, the result of \texttt{SpMM\_Mean} is fed to GCN's transformation consisting of GEMM (having Weight) and ReLU C-operations (\circleNum{3}/\circleNum{4}).
The final output should be \texttt{Result} in this DFG. Note that, ReLU is a function of MLPs, which prevents the exponential growth in the computation and vanishing gradient issue \cite{karlik2011performance}.
The user can write this DFG using our computation graph library as shown in Figure \ref{fig:dfg}.
It declares \texttt{Batch} and \texttt{Weight} by calling \texttt{CreateIn()} (lines 2$\sim$3). \texttt{BatchPre} (\circleNum{1}), \texttt{SpMM\_Mean} (\circleNum{2}), \texttt{GEMM} (\circleNum{3}), and \texttt{ReLU} (\circleNum{4}) are defined through \texttt{CreateOp()}, which are listed in lines 4$\sim$7.

GraphRunner then sorts the calling sequence of CSSD library interfaces and generates a markup file as shown in Figure \ref{fig:graphrunner3}.
This DFG final file includes a list of nodes, each defining its node sequence number, C-operation name, where the input(s) come from, and what the output(s) are.
For example, the third node is \texttt{GEMM} (\texttt{3: "GEMM"}), and its inputs come from the second node's first output (\texttt{2\_0}) as well as input node, \texttt{Weight} (\texttt{in=\{"2\_0", "Weight"\}}).
This node generates one output only (\texttt{out=\{"3\_0"\}}).

\noindent \textbf{Execution of DFG.}
The host can run CSSD with the programmed GNN by downloading the corresponding DFG and a given batch through \texttt{Run()} RPC.
As shown in Figure \ref{fig:graphrunner4}, GraphRunner's engine visits each node of the DFG and checks the node's C-operation name.
For each node, the engine finds the set of C-kernels (matched with the C-operation name) by checking the operation table.
It then refers to the device table and selects the appropriate implementation among the retrieved C-kernels based on the device priority, assigned by \texttt{RegisterDevice()}.
The engine de-refers the C-kernel's address pointer and calls it by passing the C-kernel's parameters, which can also be checked up with offloaded DFG's node information (e.g., \texttt{in=\{\dots\}}).
Note that, GraphRunner's engine performs these dynamic binding and kernel execution for all the nodes of DFG per GNN inference.

\begin{figure}[]
  \addtocounter{figure}{-2}
  \centering
  \includegraphics[width=1\linewidth,bb=0 5 247 91]{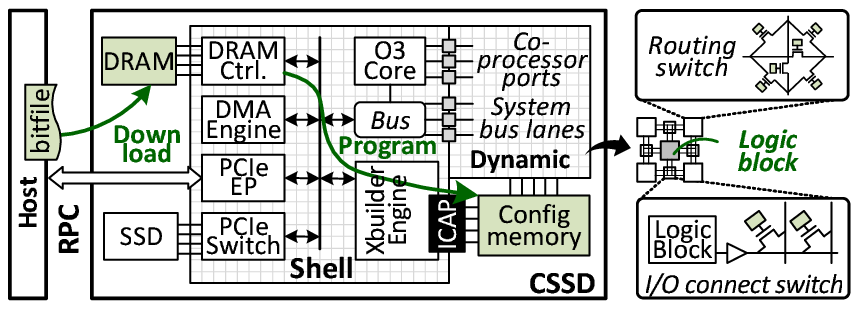}
  \vspace{-22pt} \caption{Reconfigurable hardware.\label{fig:reconfig}}
  \vspace{-20pt}
  \addtocounter{figure}{2}
\end{figure}

\subsection{Managing Reconfigurable Hardware}
\label{subsec:reconfig}
As shown in Figure \ref{fig:reconfig}, XBuilder provides static logic at Shell logic that includes an out-of-order core, a DRAM controller, DMA engines, and a PCIe switch.
This static logic is connected to User (dynamic) logic through a co-processor port such as RoCC \cite{rocketchip} and system bus (e.g., TileLink \cite{tilelink}).

XBuilder exposes the boundary position of the FPGA logic die in the form of a design checkpoint file \cite{dcp,ug947}.
The boundary is wire circuits that separate Shell and User logic, called \emph{partition pin}.
Since the static logic is fixed at the design time, we place the maximum number of co-processor ports and system bus lanes to the partition pin, which can be united with the hardware components fabricated in Shell logic.
In addition, we locate an XBuilder hardware engine in Shell, which includes the internal configuration access port (ICAP \cite{ug974,ug570}).
Note that, FPGA logic blocks are connected by many wires of routing switches and input/output connection switches.
Since the switches maintain the status of connection in built-in FPGA memory, called \emph{configuration memory}, we can reconfigure the FPGA hardware by reprogramming the connection states on the configuration memory.
As the configuration memory should be protected from anonymous accesses, FPGA only allows the users to reprogram the configuration memory only through the primitive hardware port, ICAP.
The users can simply call HolisticGNN's RPC interface, \texttt{Program} with their own hardware (partial) bitfile to reconfigure User logic.
XBuilder copies the bitfile into CSSD's FPGA internal DRAM first, and then, it reconfigures User logic by programming the logic using the bitfile via ICAP.
While User logic is being reconfigured, it would unfortunately be possible to make the static logic of Shell unable to operate appropriately.
Thus, XBuilder ties the partition pin's wires (including a system bus) by using DFX decoupler IP \cite{ug947,pg375} and makes User logic programming separate from the working logic of Shell.
In default, XBuilder implements Shell by locating an out-of-core processor and PCIe/memory subsystems that run GraphRunner and GraphStore.
Figure \ref{fig:eval_floorplan} shows three example implementation views of our Shell and User logic.
Shell logic locates an out-of-core processor and PCIe/memory subsystems that run GraphRunner and GraphStore.
In this example, we program an open-source RISC-V CPU, vector processor, and systolic array.
We will explain details of example implementations in Section \ref{sec:evaluation}.





\section{Evaluation}
\label{sec:evaluation}
\noindent \textbf{Prototypes.}
While CSSD is officially released in storage communities \cite{smartssd,scaleflux,newport}, there is no commercially available device yet.
We thus prototype a customized CSSD that employs a 14$nm$ 730MHz FPGA chip \cite{virtexup,ds923}, 16GB DDR4-2400 DRAM \cite{MTA18ASF2G72PZ}, and a 4TB high-performance SSD \cite{intelp4600} together within the same PCIe 3.0$\times$4 subsystem \cite{pcie3} as shown in Figure \ref{fig:hostpc}.
We prepare three sets of hardware accelerators for XBuilder's User logic; a multi-core processor (\emph{Octa-HGNN}), large systolic array processors (\emph{Lsap-HGNN}), and a heterogeneous accelerator having a vector processor and a systolic array (\emph{Hetero-HGNN}), as shown in Figure \ref{fig:eval_floorplan}.
Octa-HGNN employs eight out-of-order (O3) cores and performs all GNN processing using multi-threaded software.
Each O3 core is implemented based on open-source RISC-V \cite{zhaosonicboom,chipyard} having 160KB L1 and 1MB L2 caches.
For Lsap-HGNN and Hetero-HGNN, we modify an open-source SIMD (Hwacha \cite{lee2015hwacha}) and systolic architecture (Gemmini \cite{genc2021gemmini}).
In our evaluation, SIMD employs four vector units, and the systolic architecture employs 64 floating-point PEs with 128KB scratchpad memory.
Note that, all these prototypes use the same software part of HolisticGNN (GraphStore, GraphRunner, and XBuilder) as it can handle the end-to-end GNN services over DFG.

\noindent \textbf{GPU-acceleration and testbed.}
For a fair performance comparison, we also prepare two high-performance GPUs, \emph{GTX 1060} and \emph{RTX 3090}.
While GTX 1060's 10 streaming multiprocessors (SMs) operate at 1.8GHz with 6GB DRAM, RTX 3090 employs 82 SMs working at 1.7 GHz with 24GB DRAM.
To enable GNN services, we use deep graph library (DGL) 0.6.1 \cite{wang2019dgl} and TensorFlow 2.4.0 \cite{osdi16tensorflow}, which use CUDA 11.2 and cuDNN 8.2 for GPU acceleration.
DGL accesses the underlying SSD via the XFS file system to pre-/processing graphs.
The testbed uses a 2.2GHz 12-core processor with DDR4-2666 64GB DRAM and a 4TB SSD (same with the device that we used for CSSD prototype), and connect all GPUs and our CSSD prototype.
The detailed information of our real evaluation system is shown in Table \ref{tbl:testbeds}.

\begin{figure}
  \centering
  \captionsetup[figure]{aboveskip=-15pt,belowskip=-15pt}
  \captionsetup[subfigure]{aboveskip=0pt}
  \begin{subfigure}[b]{.72\linewidth}
    \includegraphics[width=1\linewidth]{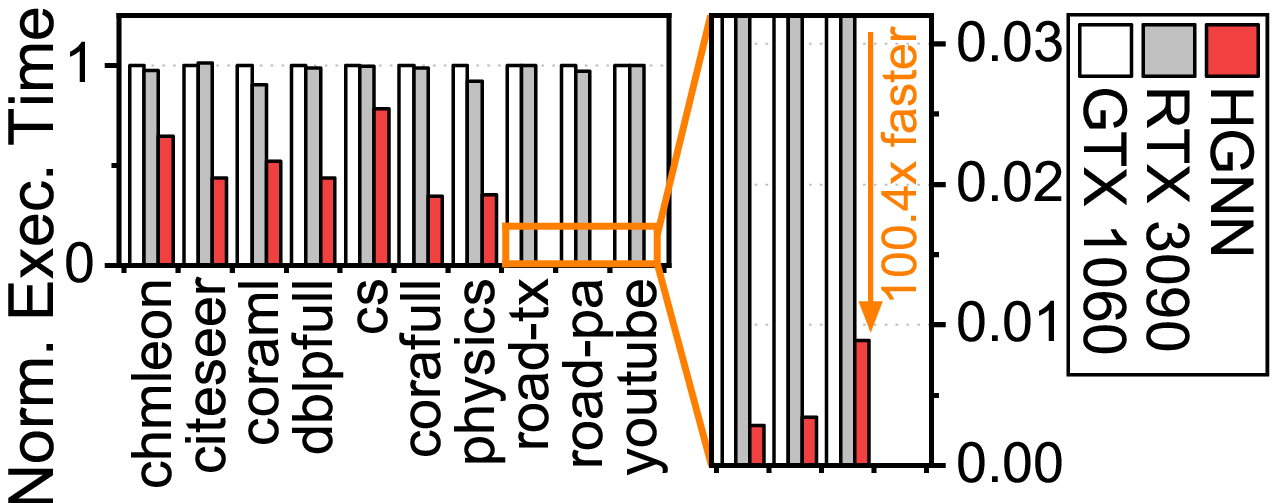}
    \caption{Normalized End-to-end latency.}
    \label{fig:eval_e2e_norm}
  \end{subfigure}
  \begin{subfigure}[b]{.27\linewidth}
    \centering
    \renewcommand{\arraystretch}{.8}
    \resizebox{\linewidth}{!}{
      \large
      \begin{tabular}{lr}
        \toprule
        \textbf{Legend} & \textbf{GTX 1060} \\ \midrule
        chmleon         &    140 ms        \\
        citeseer        &    162 ms        \\
        coraml          &    166 ms        \\
        dblpfull        &    323 ms        \\
        cs              &    618 ms        \\
        corafull        &   1233 ms        \\
        physics         &   2335 ms        \\
        road-tx         & 426732 ms        \\
        road-pa         & 332391 ms        \\
        youtube         & 341035 ms        \\ \bottomrule
        \end{tabular}
    }
    \caption{Latency.}
    \label{tbl:eval_e2e_1060}
  \end{subfigure}
  \vspace{-22pt}
  \caption{End-to-end latency comparison.}
  \label{fig:eval_e2e_gcn}
  \vspace{-19pt}
\end{figure}

\noindent \textbf{GNN models and graph datasets.}
We implement three popular GNN models, GCN \cite{kipf2016semi}, GIN \cite{xu2018powerful}, and NGCF \cite{wang2019neural}, for both GPUs and CSSD.
We also select 14 real-graph datasets (workloads) from LBC \cite{lbc}, MUSAE \cite{rozemberczki2021multi}, and SNAP \cite{leskovec2016snap}.
Since the workloads coming from SNAP \cite{leskovec2016snap} do not provide the features, we generate the features based on the feature length that the prior work (pinSAGE \cite{ying2018graph}) uses (4K).
The important characteristics of our graph datasets and workloads are described in Table \ref{tbl:dataset}.
Note that, the workloads that we listed in Table \ref{tbl:testbeds} is sorted in ascending order of their graph size.
For better understanding, we summarize the characteristics for graph before batch preprocessing (\emph{Original Graph}) and after batch preprocessing (\emph{Sampled Graph}).

\subsection{End-to-end Performance Comparisons}
\noindent \textbf{Overall latency.}
Figure \ref{fig:eval_e2e_norm} compares the end-to-end inference latency of GTX 1060, RTX 3090, and our HolisticGNN (\emph{HGNN}) using the heterogeneous hardware acceleration.
For better understanding, the end-to-end latency of RTX 3090 and HGNN is normalized to that of GTX 1060.
The actual latency value of GTX 1060 is also listed in Table \ref{tbl:eval_e2e_1060}.
We use GCN as representative of GNN models for the end-to-end performance analysis; since we observed that the pure inference computing latency only accounts for 1.8\% of total latency, the performance difference among the GNN models that we tested are negligible in this analysis (<1.1\%).
We will show the detailed inference latency analysis on the different GNN models in Section \ref{subsec:inference}.

One can observe from the figure that HGNN shows 7.1$\times$ and 7.0$\times$ shorter end-to-end latency compared to GTX 1060 and RTX 3090 across all the graph datasets except for \texttt{road-ca}, \texttt{wikitalk}, and \texttt{ljournal}.
Note that both GTX 1060 and RTX 3090 cannot execute such large-scale graphs due to the out-of-memory issue, and thus, we exclude them in this comparison.
Specifically, for the small graphs (<1M edges), HGNN outperforms GPUs by 1.69$\times$, on average.
This performance superiority of HGNN becomes higher when we infer features on large-scale graphs (>3M edges), which makes HGNN 201.4$\times$ faster than GTX 1060 and RTX 3090, on average.
Even though the operating frequency and computing power of GTX 1060 and RTX 3090 are much better than HGNN, most of data preprocessing for both graphs and batches are performed by the host, and its computation is involved in storage accesses.
This in turn makes the end-to-end inference latency longer.
In contrast, HGNN can preprocess graphs in parallel with the graph updates and prepare sampled graphs/embeddings directly from the internal SSD, which can successfully reduce the overhead imposed by preprocessing and storage accesses.
We will dig deeper into the performance impact of preprocessing/storage (GraphStore) and hardware accelerations (XBuilder) shortly.


\begin{figure}
  \includegraphics[width=1\linewidth]{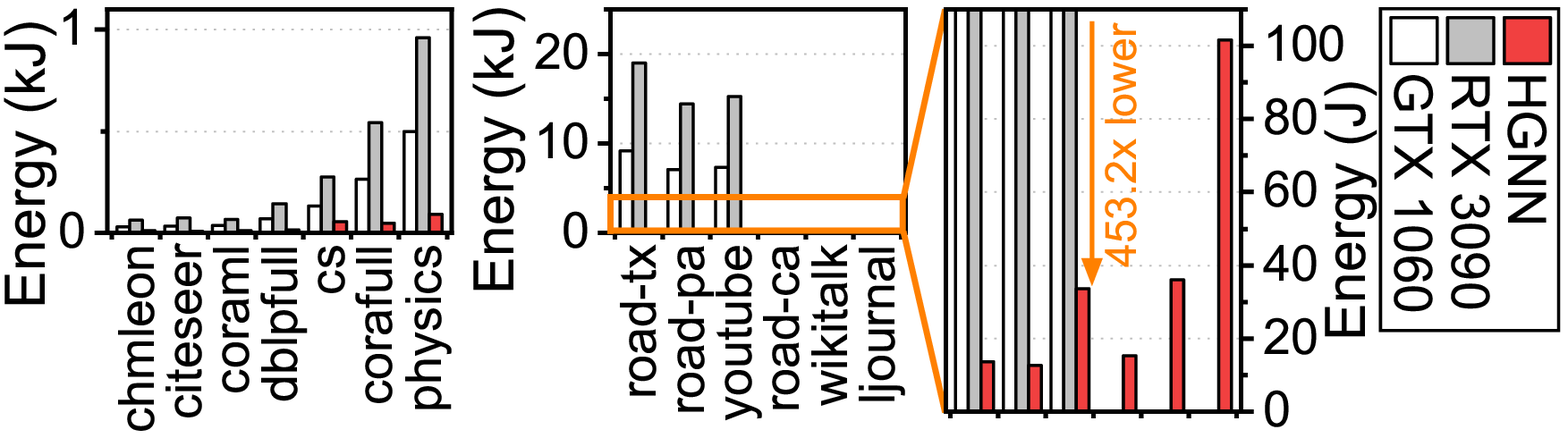}
  \vspace{-24pt}
  \caption{Estimated energy consumption comparison.}
  \label{fig:eval_power_gcn}
  \vspace{-20pt}
\end{figure}

\begin{figure*}
  \centering
  \captionsetup[figure]{aboveskip=-15pt,belowskip=-15pt}
  \captionsetup[subfigure]{aboveskip=0pt}
  \includegraphics[width=1\linewidth,bb=0 30 1023 159]{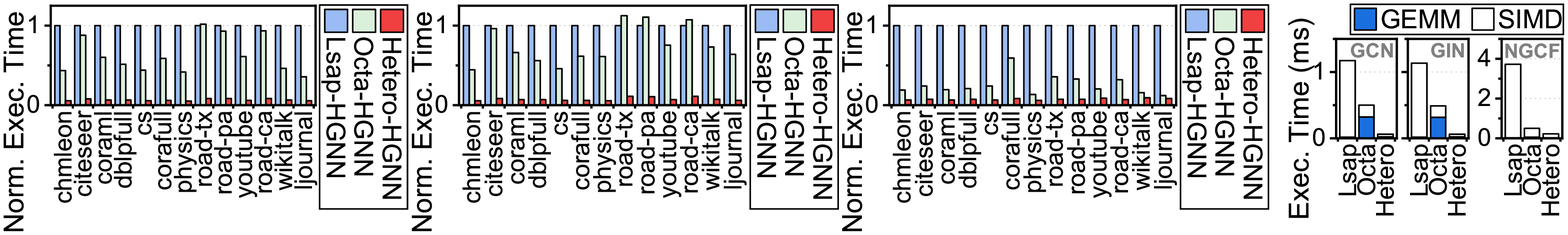}
  \begin{minipage}{.81\linewidth}
    \begin{subfigure}{.33\linewidth}
      \caption{GCN.}
    \end{subfigure}
    \begin{subfigure}{.33\linewidth}
      \caption{GIN.}
    \end{subfigure}
    \begin{subfigure}{.33\linewidth}
      \caption{NGCF.}
      \label{fig:eval_reconfig_ngcf}
    \end{subfigure}
    \vspace{-20pt}
    \caption{Inference latency comparison.}
    \label{fig:eval_reconfig}
  \end{minipage}
  \begin{minipage}{.18\linewidth}
    \vspace{15pt}
    \caption{Breakdown.}
    \label{fig:eval_reconfig_break}
  \end{minipage}
  \vspace{-25pt}
\end{figure*}

\noindent \textbf{Energy consumption.}
Figure \ref{fig:eval_power_gcn} analyzes the energy consumption behaviors of all three devices we tested.
Even though GTX 1060 and RTX 3090 show similar end-to-end latency behaviors in the previous analysis, RTX 3090 consumes energy 2.04$\times$ more than what GTX 1060 needs because it has 8.2$\times$ and 4$\times$ more SMs and DRAM, respectively.
In contrast, HGNN exhibits 33.2$\times$ and 16.3$\times$ better energy consumption behaviors compared to RTX 3090 and GTX 1060, on average, respectively.
Note that, HGNN processes large-scale graphs by consuming as high as 453.2$\times$ less energy than the GPUs we tested.
This is because, in addition to the latency reduction of HGNN, our CSSD consumes only 111 Watts at the system-level thanks to the low-power computing of FPGA (16.3 Watts).
This makes HGNN much more promising on GNN computing compared to GPU-based acceleration approaches.
Note that, RTX 3090 and GTX 1060 consume 214 and 447 Watts at the system-level, respectively.

\subsection{Pure Inference Acceleration Comparison}
\label{subsec:inference}
Figure \ref{fig:eval_reconfig} shows the pure inference performance of Hetero-HGNN and Octa-HGNN, normalized to Lsap-HGNN; before analyzing the end-to-end service performance, we first compare HolisticGNN itself different User logic here.

One can observe from this figure that, even though systolic arrays are well optimized for conventional DL such as CNN and RNN, Lsap-HGNN exhibits much worse performance than software-only approach.
For all the graph datasets that we tested, Octa-HGNN exhibits shorter GNN inference latency compared to Lsap-HGNN by 2.17$\times$, on average. 
This is crystal clear evidence that the conventional DL hardware acceleration is not well harmonized with GNN inference services.
Since the computation of aggregation is involved in traversing the graph data, the systolic arrays (bigger than any hardware logic that we tested) cannot unfortunately accelerate the inference latency.
In contrast, Octa-HGNN processes the aggregation (including transformation) over multi-processing with many cores in User logic.
As shown in Figure \ref{fig:eval_reconfig_ngcf}, this phenomenon is more notable on the inference services with NGCF (4.35$\times$ faster than Lsap-HGNN) because NGCF has more heavier aggregation (similarity-aware and element-wise product explained in Section \ref{subsec:gnn}).

However, the performance of Octa-HGNN is also limited because matrix computation on dense embeddings (GEMM) is not well accelerated by its general cores.
In contrast, Hetero-HGNN has both SIMD and systolic array units, which are selectively executed considering the input C-kernel, such that Hetero-HGNN shortens the inference latency of Octa-HGNN and Lsap-HGNN by 6.52$\times$ and 14.2$\times$, on average, respectively.
Figure \ref{fig:eval_reconfig_break} decomposes the inference latency of three HGNN that we tested into SIMD and GEMM for a representative workload, physics.
As shown in figure, Lsap-HGNN mostly exhibits GEMM as its systolic arrays accelerate the transformation well, but its performance slows down due to a large portion of SIMD.
The latency of Octa-HGNN suffers from GEMM computation, which accounts for 34.8\% of its inference latency, on average.
As Hetero-HGNN can accelerate both SIMD and GEMM, it successfully shortens the aggregation and transformation for all GNN models that we tested.
This is the reason why we evaluate the end-to-end GNN latency using Hetero-HGNN as a default hardware acceleration engine in the previous section.


\begin{figure*}
  \centering
  \captionsetup[figure]{aboveskip=-15pt,belowskip=-15pt}
  \captionsetup[subfigure]{aboveskip=0pt}
  \includegraphics[width=1\linewidth]{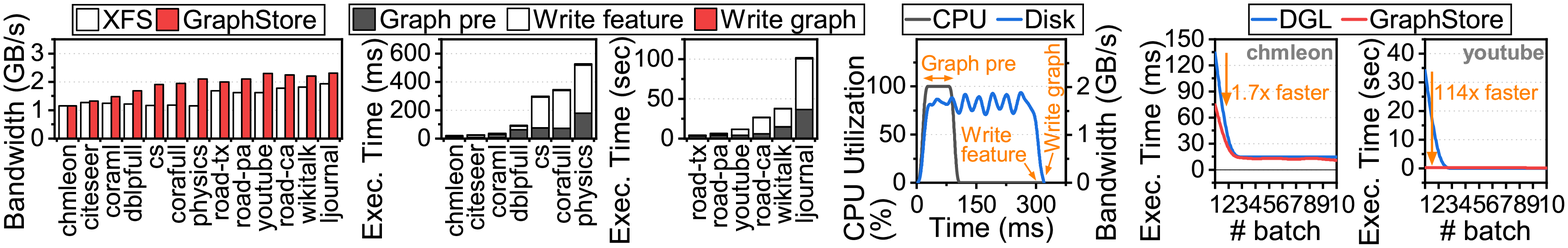}
  \begin{minipage}{.72\linewidth}
    \begin{subfigure}{.3\linewidth}
      \caption{Peak write bandwidth.}
      \label{fig:eval_init_bw}
    \end{subfigure}
    \begin{subfigure}{.44\linewidth}
      \caption{Latency breakdown.}
      \label{fig:eval_init}
    \end{subfigure}
    \begin{subfigure}{.24\linewidth}
      \caption{Timeline of \texttt{cs}.}
      \label{fig:eval_init_timeline}
    \end{subfigure}
    \vspace{-10pt}
    \caption{Performance analysis of GraphStore bulk operations.}
  \end{minipage}
  \begin{minipage}{.27\linewidth}
    \vspace{5pt}
    \caption{Multiple batches.}
    \label{fig:eval_multi_batch}
  \end{minipage}
  \vspace{-20pt}
\end{figure*}


\subsection{Performance Analysis on GraphStore}
\noindent \textbf{Bulk operations.}
Figures \ref{fig:eval_init_bw} and \ref{fig:eval_init} show the bandwidth and latency of GraphStore's bulk operations.
While the GPU-enabled host system writes the edge array and corresponding embeddings to the underlying SSD through its storage stack, GraphStore directly writes the data to internal storage without any storage stack involvement.
This does not even exhibit data copies between page caches and user-level buffers, which in turn makes GraphStore exposes performance closer to what the target SSD actually provides.
As a result, GraphStore shows 1.3$\times$ better bandwidth on graph updates compared to conventional storage stack (Figure \ref{fig:eval_init_bw}).
More importantly, GraphStore hides the graph preprocessing overhead imposed by converting the input dataset to the corresponding adjacency list with the update times of heavy embeddings.
We also show how much the embedding update (\texttt{Write feature}) can hide the latency of graph preprocessing (\texttt{Graph pre}) in Figure \ref{fig:eval_init}.
Since \texttt{Write feature} in the figure only shows the times longer than \texttt{Graph pre}, we can observe that GraphStore can make \texttt{Graph pre} completely invisible to users.
For better understanding, we also perform a time series analysis of \texttt{cs} as an example of other workloads, and the results are shown Figure \ref{fig:eval_init_timeline}.
The figure shows the dynamic bandwidth in addition to the per-task utilization of Shell's simple core.
As shown in the figure, GraphStore starts the preprocessing as soon as it begins to write the embeddings to the internal SSD.
\texttt{Graph pre} finishes at 100$ms$ while \texttt{Write feature} ends at 300$ms$.
Thus, \texttt{Write feature} is performed with the best performance of the internal SSD (around 2GB/s).
Note that, even though writing the adjacency list \texttt{Write graph} is performed right after \texttt{Write feature} (Figure \ref{fig:eval_init}), it is almost invisible to users (Figure \ref{fig:eval_init_timeline}) as the size of graph is much smaller than the corresponding embeddings (357.1$\times$, on average).

\noindent \textbf{Batch preprocessing (Get).}
Figure \ref{fig:eval_multi_batch} shows batch preprocessing, which is the only task to read (sub)graphs from the storage in the end-to-end viewpoint; node sampling and embedding lookup use \texttt{GetNeighbor()} and \texttt{GetEmbed()}, respectively.
In this evaluation, we compare batch preprocessing performance of GPU-enabled host and CSSD using \texttt{chmleon} and \texttt{youtube} each being representative of small and large graphs.
For the earliest batch preprocessing, GraphStore performs batch preprocessing 1.7$\times$ (\texttt{chmleon}) and 114.5$\times$ (\texttt{youtube}) faster than that of the GPU-enabled host, respectively.
Even though GraphStore is working at a lower frequency (3$\times$ than the host CPU), \texttt{GetNeighbor()} and \texttt{GetEmbed()} are much faster because the graph data has been already converted into an adjacency list at the graph update phase.
In contrast, the host needs to process the graph data at the first batch, such that node sampling and embedding lookup can find out the appropriate targets.
After the first batch, both cases, mostly accessing the neighbors and the corresponding embeddings are processed in memory thereby showing sustainable performance.
Note that, even though we showed the batch preprocessing performance for only \texttt{chmleon} and \texttt{youtube} (due to the page limit), this performance trend is observed across all the workloads that we tested.

\begin{figure}[b]
  \vspace{-15pt}
  \centering
  \includegraphics[width=1\linewidth]{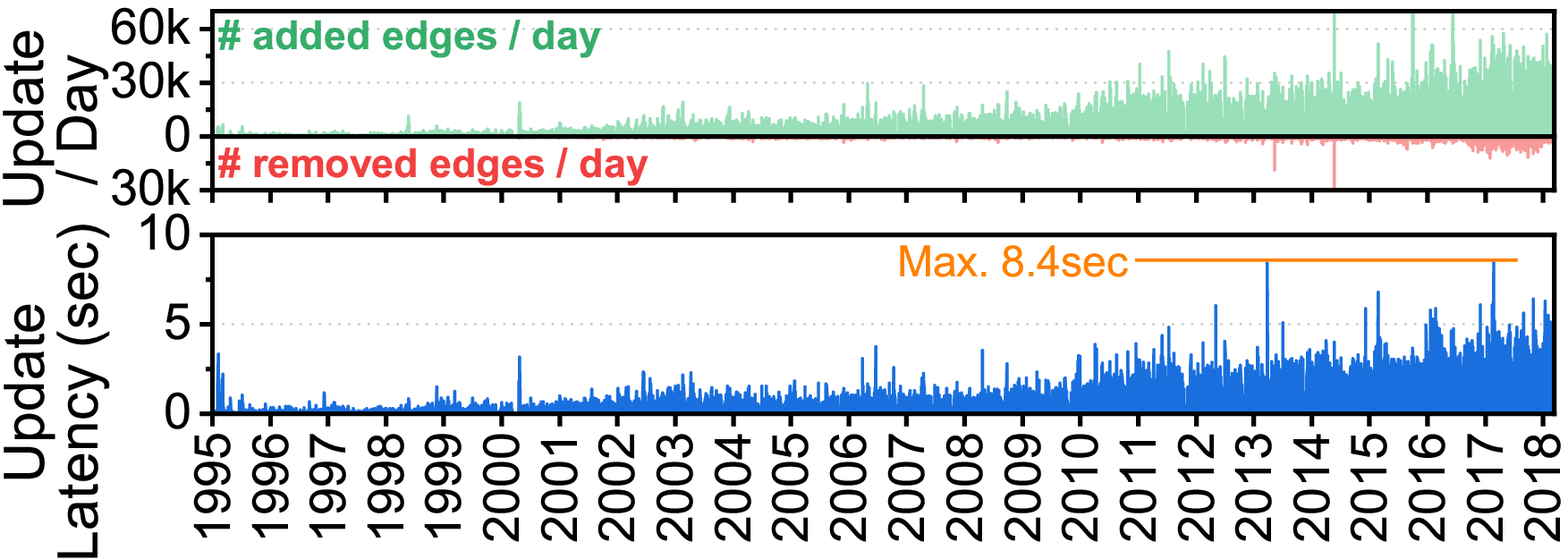}
  \vspace{-21pt}
  \caption{GraphStore update performance.}
  \label{fig:eval_mutable}
  \vspace{-5pt}
\end{figure}

\noindent \textbf{Mutable graph support (Unit operations).}
Since there is no publicly available dataset for mutable graph support, we evaluate the unit operations (requested by the host to CSSD) by processing historical DBLP datasets \cite{hoffmann_oliver_2019_3051910}.
The top of Figure \ref{fig:eval_mutable} shows the number of per-day add and delete operations for the past 23 years (1995$\sim$2018), and its bottom shows the corresponding per-day (accumulated) latency of GraphStore.
The workload adds 365 new nodes and 8.8K new edges into GraphStore, and deletes 16 nodes and 713 edges per day, on average.
As shown in Figure, GraphStore exhibits 970$ms$ for per-day updates, on average, and the accumulated latency in the worst case of GraphStore is just 8.4 sec, which takes reasonably short in the workload execution time (0.01\%).

\section{Related Work and Discussion}
\label{sec:relatedwork}
There are many studies for in-storage processing (ISP) \cite{jo2016yoursql,koo2017summarizer,seshadri2014willow,gu2016biscuit,lee2014accelerating,quero2015self}, including DL accelerating approaches such as \cite{wilkening2021recssd, liang2019cognitive,mailthody2019deepstore}. All these studies successfully brought significant performance benefits by removing data transferring overhead.
However, these in-storage, smart storage approaches require fully integrating their computations into an SSD, which unfortunately makes the data processing deeply coupled with flash firmware and limited to a specific computing environment that the storage vendor/device provides.
These approaches also use a thin storage interface to communicate with the host and the underlying SSD, which require a significant modification of application interface management.
More importantly, all they are infeasible to accelerate GNN computing, which contains both graph-natured processing and DL-like dense computing operations.

On the other hand, architectural research \cite{liang2020engn,yan2020hygcn,auten2020hardware} focuses on accelerating GNN core over a fixed hardware design such as vector units and systolic processors. While this simulation-based achieves the great performance benefit on GNN inference, they are ignorant of performance-critical components such as graph preprocessing and node sampling. These simulation-based studies also assume that their accelerator can have tens of hundreds of preprocessing elements (PEs), which may not be feasible to integrate into CSSD because of the hardware area limit. In contrast, HolisticGNN accelerates GNN-related tasks from the beginning to the end near storage, and its real system implementation only contains 64 PEs for the GNN inference acceleration.

Lastly, there are FPGA approaches to deep learning accelerations \cite{guo2017angel, zhang2017frequency, wu2017high}. Angel-Eye \cite{guo2017angel} quantizes data to compress the original network to a fixed-point form and decrease the bit width of computational parts. A frequency-domain hybrid accelerator \cite{zhang2017frequency} applies discrete Fourier transformation methods to reduce the number of multiplications of convolutions. On the other hand, a reconfigurable processing array design \cite{wu2017high} tries to increase the operating frequency of any target FPGA in order to build a high throughput reconfigurable processing array.
These studies are unfortunately not feasible to capture the GNN acceleration, and cannot eliminate the preprocessing overhead imposed by graph-natured complex computing near storage.
Note that, it would be possible to use cross-platform abstraction platforms, such as OpenCL \cite{stone2010opencl} or SYCL \cite{reyes2016sycl}, rather than using RPC.
OpenCL/SYCL is excellent for managing all hardware details at a very low-level, but they can bump up the complexity of what users need to control. For example, users should know all heterogeneities of reconfigurable hardware for the end-to-end GNN acceleration and handle CSSD’s memory space over OpenCL/SYCL.

\section{Conclusion}
\label{sec:conclusion}
We propose HolisticGNN that provides an easy-to-use, near-storage inference infrastructure for fast, energy-efficient GNN processing. To achieve the best end-to-end latency and high energy efficiency, HolisticGNN allows users to implement various GNN algorithms close to the data source and execute them directly near storage in a holistic manner.
Our empirical evaluations show that the inference time of HolisticGNN outperforms GNN inference services using high-performance modern GPUs by 7.1$\times$ while reducing energy consumption by 33.2$\times$, on average.

\section*{Acknowledgements}
This research is supported by Samsung Research Funding \& Incubation Center of Samsung Electronics (SRFC-IT2101-04).
This work is protected by one or more patents, and Myoungsoo Jung is the corresponding author.
The authors would like to thank the anonymous reviewers for their comments and suggestions.
The authors also thank Raju Rangaswami for shepherding this paper.


\bibliographystyle{plain}
\bibliography{ref}

\end{document}